\documentclass{LMCS}

\def\dOi{10(4:13)2014}
\lmcsheading%
{\dOi}
{1--27}
{}
{}
{Apr.~10, 2013}
{Dec.~18, 2014}
{}

\ACMCCS{[{\bf Theory of computation}]: Models of computation; Logic;
  Formal languages and automata theory} 

\usepackage{enumerate}
\usepackage{hyperref}

\usepackage{ifthen}
\usepackage{rotating}
\usepackage{graphicx}
\usepackage{amssymb}
\usepackage{xspace}
\usepackage{subfig}
\usepackage{tikz}
\usetikzlibrary{automata}
\usepackage{multirow}

\newcommand{\hide}[1]{}
\newcommand{\buchi}{{B\"uchi}\xspace}
\renewcommand{\inf}{\mathit{inf}}

\newcommand{\f}{\mathcal{F}}
\newcommand{\g}{\mathcal{G}}

\newcommand{\rulename}[1]{\texttt{#1}}

\newcommand{\projq}[1]{{#1}_{\downarrow Q}}
\newcommand{\projd}[1]{{#1}_{\downarrow D}}
\newcommand{\merge}{\mathit{merge}}
\newcommand{\ramsey}{\textsc{Ramsey}\xspace}
\newcommand{\ramseyp}{\textsc{Ramsey}+P\xspace}
\newcommand{\ramseya}{\textsc{Ramsey}+A\xspace}
\newcommand{\ramseym}{\textsc{Ramsey}+m\xspace}
\newcommand{\ramseyam}{\textsc{Ramsey}+Am\xspace}
\newcommand{\ramseypam}{\textsc{Ramsey}+PAm\xspace}
\newcommand{\SP}{\textsc{SP}\xspace}
\newcommand{\SPp}{\textsc{SP}+P\xspace}
\newcommand{\SPa}{\textsc{SP}+A\xspace}
\newcommand{\SPs}{\textsc{SP}+S\xspace}
\newcommand{\SPe}{\textsc{SP}+E\xspace}
\newcommand{\SPase}{\textsc{SP}+ASE\xspace}
\newcommand{\SPpase}{\textsc{SP}+PASE\xspace}
\newcommand{\piterman}{\textsc{Safra-Piterman}\xspace}

\newcommand{\pitermanase}{\textsc{Safra-Piterman}+ASE\xspace}

\newcommand{\rank}{\textsc{Rank}\xspace}
\newcommand{\rankp}{\textsc{Rank}+P\xspace}
\newcommand{\ranka}{\textsc{Rank}+A\xspace}
\newcommand{\rankpa}{\textsc{Rank}+PA\xspace}
\newcommand{\slice}{\textsc{Slice}\xspace}
\newcommand{\slicep}{\textsc{Slice}+P\xspace}
\newcommand{\slicea}{\textsc{Slice}+A\xspace}
\newcommand{\sliced}{\textsc{Slice}+D\xspace}
\newcommand{\slicer}{\textsc{Slice}+R\xspace}
\newcommand{\slicem}{\textsc{Slice}+M\xspace}
\newcommand{\sliceadrm}{\textsc{Slice}+ADRM\xspace}
\newcommand{\slicepadrm}{\textsc{Slice}+PADRM\xspace}

\newcommand{\set}[1]{\mathcal{#1}}

\newcommand{\todo}[1]{(\textbf{TODO}: #1)}

\begin{document}

\title[State of B\"uchi Complementation]{State of B\"uchi Complementation\rsuper*}

\author[M.-H.~Tsai]{Ming-Hsien Tsai\rsuper a}
\address{{\lsuper{a,d}}National Taiwan University}
\email{mhtsai208@gmail.com, tsay@im.ntu.edu.tw}
\thanks{{\lsuper a}Work supported in part by the National Science Council, Taiwan
  (R.O.C.) under grants NSC97-2221-E-002-074-MY3 and
  NSC102-2221-E-002-090, by NSF grants CCF-0613889, ANI-0216467,
  CCF-0728882, and OISE-0913807, by BSF grant 9800096, and by gift
  from Intel.}

\author[S.~Fogarty]{Seth Fogarty\rsuper b}
\address{{\lsuper b}Trinity University}
\email{sfogarty@trinity.edu}

\author[M.Y.~Vardi]{Moshe Y.~Vardi\rsuper c}
\address{{\lsuper c}Rice University}
\email{vardi@cs.rice.edu}

\author[Y.-K.~Tsay]{Yih-Kuen Tsay\rsuper d}



\titlecomment{{\lsuper*}A preliminary version of this paper appeared in the
  Proceedings of the 15th International Conference on Implementation
  and Application of Automata (CIAA), Lecture Notes in Computer
  Science 6482, Springer-Verlag, 2011, pp. 261-271.}


\begin{abstract}
Complementation of \buchi automata has been studied for over five
decades since the formalism was introduced in 1960.
Known complementation constructions can be classified into
Ramsey-based, determinization-based, rank-based, and slice-based
approaches.
Regarding the performance of these approaches, there have been several
complexity analyses but very few experimental results.
What especially lacks is a comparative experiment on all of the four
approaches to see how they perform in practice.
In this paper, we review the four approaches, propose several optimization
heuristics, and perform comparative experimentation on four representative
constructions that are considered the most efficient in each approach.
The experimental results show that (1) the determinization-based
Safra-Piterman construction outperforms the other three in producing smaller
complements and finishing more tasks in the allocated time and (2) the
proposed heuristics substantially improve the Safra-Piterman
and the slice-based constructions.
\end{abstract}

\maketitle\vfill

\section{Introduction}

\buchi automata are nondeterministic finite automata on infinite words.
They recognize $\omega$-regular languages and are closed under Boolean
operations, namely union, intersection, and complementation.
The formalism was first proposed and studied by \buchi in 1960 as part
of a decision procedure for second-order logic~\cite{buchi:decision}. 
Complementation of \buchi automata is significantly more difficult than that
of nondeterministic finite automata on finite words.
Given a nondeterministic finite automaton on finite words with $n$ states,
complementation yields an automaton with $2^n$ states through the subset
construction~\cite{RS59}.
Indeed, the subset construction is insufficient for the complementation of
nondeterministic \buchi automata.
In fact, Michel showed in 1988 that the blow-up of \buchi complementation is at
least $n!$ (approximately $(n/e)^n$ or $(0.36n)^n)$, which is much higher than
$2^n$~\cite{michel:complementation}.
This lower bound was eventually sharpened by Yan to
$(0.76n)^n$~\cite{yan:lower}, which was matched by an upper bound by
Schewe~\cite{schewe:buchi}.

There are several applications of \buchi complementation in formal verification.
For example, whether a system satisfies a property can be verified by checking
if the intersection of the system automaton and the complement of the property
automaton is empty~\cite{Var96}.
Another example is that the correctness of an LTL translation algorithm can be
tested with a reference algorithm as done in the development of the
GOAL tool~\cite{tsay:goal,tsai:goal}\footnote{With the help of
  complementation, implementations of translation algorithms in
  GOAL have been tested against formulae in
  Spec Patterns~\cite{spec:patterns} and randomly generated
  formulae.}.
Moreover, \buchi complementation also involves in the translation of
QPTL~\cite{kesten:complete} and ETL~\cite{wolper:temporal} formulae.
Both QPTL and ETL are more expressive than LTL.
Although recently many works have focused on universality and
containment testing without going explicitly through
complementation~\cite{fogarty:buchi,fogarty:efficient,doyen:antichains}, it is
still unavoidable in some cases~\cite{KV05c}.

There have been quite a few complementation constructions, which can
be classified into four approaches: Ramsey-based
approach~\cite{buchi:decision,sistla:complementation},
determinization-based
approach~\cite{safra:complexity,muller:simulating,althoff:observations,piterman:buchi}, 
rank-based
approach~\cite{thomas:complementation,kupferman:weak,klarlund:progress}, 
and slice-based
approach~\cite{kahler:complementation,vardi:automata07}.
The second approach is a deterministic approach while the last two are
nondeterministic approaches.
The first three approaches were reviewed in~\cite{vardi:buchi}.
Due to the high complexity of \buchi complementaton, optimization
heuristics are critical to good
performance~\cite{gurumurthy:complementing,friedgut:buchi,schewe:buchi,karmarkar:minimal,klein:experiments}.
However, with much recent emphasis shifted to universality and
containment, empirical studies of \buchi complementation have been
scarce~\cite{klein:experiments,gurumurthy:complementing,karmarkar:minimal,tsay:goalextended}
in contrast with the rich theoretical developments.
A comprehensive empirical study would allow us to evaluate the performance
of these complementation approaches that has so far been characterized
only by theoretical bounds.

In this paper, we review the four complementation approaches and perform 
comparative experimentation on four selected constructions that we
consider the best in each approach.
All the four constructions have been implemented in
GOAL~\cite{tsay:goal,tsai:goal}.
Although one might expect that the nondeterministic
approaches would be better than the deterministic approach because of
better worse-case bounds, our experimental results show that the
deterministic construction is the best for complementation in average.
At the same time the Ramsey-based approach, which is competitive in
universality and containment
testing~\cite{abdulla:when,fogarty:buchi,fogarty:efficient},
performs rather poorly in our complementation experiments.
We also propose optimization heuristics for the
determinization-based construction, the rank-based construction, and the
slice-based construction.
Our experiment shows that the optimization heuristics substantially improve the
three constructions.
Overall, our work confirms the importance of experimentation and
heuristics in studying \buchi complementation, as worst-case bounds
may not be accurate indicators of performance.

The rest of this paper is organized as follows.
Some preliminaries are given in Section~\ref{sec:preliminaries}.
In Section~\ref{sec:review}, we review the four complementation
approaches.
We discuss the results of our comparative experimentation on the four
approaches in Section~\ref{sec:comparison}.
Section~\ref{sec:optimization} describes our optimization heuristics
and Section~\ref{sec:experiment} shows the improvement made by our
heuristics. 
We conclude in Section \ref{sec:conclusion}.

\section{Preliminaries}\label{sec:preliminaries}
\enlargethispage{\baselineskip}

A (\emph{nondeterministic}) $\omega$-automaton is a five tuple $(\Sigma,
Q, q_0, \delta, \f)$, where $\Sigma$ is a nonempty finite alphabet, $Q$ is
a nonempty finite set of states, $q_0 \in Q$ is the initial state, $\delta: Q
\times \Sigma \rightarrow 2^Q$ is the transition function, and $\f$ is
the acceptance condition, to be described subsequently.
The automaton is \emph{deterministic} if $|\delta(q, a)| = 1$ for all
$q \in Q$ and $a \in \Sigma$.

Let $A = (\Sigma, Q, q_0, \delta, \f)$ be an $\omega$-automaton and $w
= a_0 a_1 \cdots \in \Sigma^\omega$ an infinite word.
A \emph{run} of $A$ on $w$ is a sequence $q_0 q_1 \cdots \in Q^\omega$
satisfying $\forall i : q_{i+1} \in \delta(q_i, a_i)$.
A run is \emph{accepting} if it satisfies the acceptance condition and
a word is \emph{accepted} if there is an accepting run on it.
The \emph{language} of an $\omega$-automaton $A$, denoted by $L(A)$,
is the set of words accepted by $A$.

Let $\rho$ be a run and $\inf(\rho)$ be the set of states that occur
infinitely often in $\rho$.
Various types of $\omega$-automata can be defined by assigning
different acceptance conditions as follows. 
\begin{itemize}
\item \emph{\buchi condition}: $\f \subseteq Q$.
  $\rho$ satisfies the condition iff $\inf(\rho) \cap \f \neq
  \emptyset$, and every $q \in \f$ is called an \emph{accepting
    state}.
\item \emph{Muller condition}: $\f \subseteq 2^Q$.
  $\rho$ satisfies the condition iff there exists an $F \in \f$ such
  that $\inf(\rho) = F$.
\item \emph{Rabin condition}: $\f \subseteq 2^Q \times 2^Q$.
  $\rho$ satisfies the condition iff there exists a pair $(E, F) \in
  \f$ such that $\inf(\rho) \cap E = \emptyset$ and $\inf(\rho) \cap F
  \neq \emptyset$. 
\item \emph{Streett condition}: $\f \subseteq 2^Q \times 2^Q$.
  $\rho$ satisfies the condition iff for all pairs $(E, F) \in \f$,
  $\inf(\rho) \cap F \neq \emptyset$ implies $\inf(\rho) \cap E \neq
  \emptyset$.
\item \emph{parity condition}: $\f : Q \rightarrow \{0, 1, \ldots,
  2r\}$.
  $\rho$ satisfies the condition iff $\mathit{min}\{\f(q) \mid q \in
  \inf(\rho) \}$ is even and $\f(q)$ is called the parity of the state
  $q$.
\end{itemize}
The parity condition $\f : Q \rightarrow \{0, 1, \ldots, 2r\}$ is a
special case of the Rabin condition, refered to as \emph{Rabin chain},
$\{ (E_0, F_0), \ldots, (E_r, F_r) \}$ where $E_0 \subset F_0 \subset
E_1 \subset \cdots \subset E_r \subset F_r$, $E_i = \{q \in Q : \f(q)
< 2i\}$, and $F_i = \{q \in Q : \f(q) \leq 2i\}$.

We use a system of three-letter acronyms to denote these
$\omega$-automata.
The first letter indicates whether the automaton is
\textbf{n}ondeterministic or \textbf{d}eterministic. 
The second letter indicates whether the acceptance condition is
\textbf{B}\"uchi, \textbf{M}uller, \textbf{R}abin, \textbf{S}treett,
or \textbf{p}arity. 
The third letter is always a ``\textbf{W}'' indicating that the
automaton accepts words.
For example, NBW stands for a nondeterministic \buchi automaton and
DPW stands for a deterministic parity automaton.

Let $A$ be an $\omega$-automaton with an alphabet $\Sigma$.
$A$ is \emph{universal} iff $L(A) = \Sigma^\omega$.
A complement of $A$ is defined as an automaton that accepts exactly
the language $\Sigma^\omega - L(A)$, denoted by $\overline{L(A)}$ when
the alphabet $\Sigma$ is clear in the context.
A state is \emph{live} if it occurs in an accepting run on some word,
and is \emph{dead} otherwise.
Dead states can be discovered using a nonemptiness algorithm,
cf. \cite{Var07}, and can be pruned off without affecting the language
of the automaton.
As a complement of a universal automaton has no accepting run, only
the initial state will remain in the complement after pruning dead
states.

When focusing on \buchi conditions, the acceptance of an infinite word
$w$ by an NBW $A$ can be determined not only by the sequential runs of
$A$ on $w$ but also by an aggregated tree structure of those runs.
Depending on different ways of aggregation, different tree structures
can be defined.

The \textit{run tree} of $A$ on $w$ is a tree where a (full) branch
corresponds to a run of $A$ on $w$ and there is a corresponding branch
for each run of $A$ on $w$.
To determine whether $w$ is accepted by $A$, one needs only pay
attention to the distinction between accepting and non-accepting
states, and a run tree may be simplified or abstracted to leave just
this much detail.
The \emph{split tree} of $A$ on $w$ is a binary tree that abstracts the
run tree by grouping accepting children and nonaccepting children of a
tree node respectively into a left child and a right child of the
node.
The word $w$ is accepted by $A$ if there is a \emph{left-recurring
  branch} in the split tree of $A$ on $w$ while a branch is
left-recurring if the branch goes left infinitely many times.
Define tree width as the maximal number of nodes that are on the same
level.
Both run trees and split trees suffer the problem of unbounded tree
width, which motivates the next tree structure.
The \emph{reduced split tree} of $A$ on $w$ is a binary tree obtained
from the split tree of $A$ on $w$ by removing a state from a node
if it also occurs in a node to the left on the same level; a node
is removed if it becomes empty.
Similarly, $w$ is accepted by $A$ if there is a left-recurring branch
in the reduced split tree of $A$ on $w$.
Each a split tree or a reduced split tree can be represented by a
sequence of \emph{slices} where a slice is a sequence of sets of
states representing all nodes on a same level of the tree from left
to right.

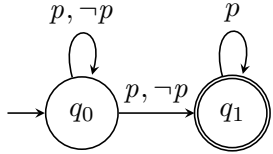
\begin{figure}[htb]
\begin{center}
\begin{tikzpicture}[
  auto, initial text=,
  semithick, >=stealth
]
  \node[state,initial](q0) at (1.6,0.72){$q_0$};
  \node[state,accepting](q1) at (3.6,0.72){$q_1$};
  \path[->]
    (q0) edge [loop above] node {$\mathit{p, \neg p}$} ()
    (q0) edge [] node {$\mathit{p, \neg p}$} (q1)
    (q1) edge [loop above] node {$p$} ()
  ;
\end{tikzpicture}
\end{center}
\caption{An NBW where the alphabet is $\{p, \neg p\}$, the initial
  state is $q_0$, and the acceptance condition is $\{q_1\}$\label{fig:nbw}}
\end{figure}

\begin{figure}[htb]
\subfloat[][the split tree of the NBW in Figure~\ref{fig:nbw} on the accepted word
$p \neg p p^\omega$]{\label{fig:split_tree} 
\begin{tikzpicture}[
  auto, initial text=,
  semithick, >=stealth,
]
  \node [] (root) at (0,0) {$\{q_0\}$};
  \node [] (l1_1) at (-0.5,-1) {$\{q_1\}$};
  \node [] (l1_2) at (0.5,-1) {$\{q_0\}$};
  \node [] (l2_2) at (0,-2) {$\{q_1\}$};
  \node [] (l2_3) at (1,-2) {$\{q_0\}$};
  \node [] (l3_2) at (-.5,-3) {$\{q_1\}$};
  \node [] (l3_3) at (.5,-3) {$\{q_1\}$};
  \node [] (l3_4) at (1.5,-3) {$\{q_0\}$};
  \node [] (l4_2) at (-1,-4) {$\{q_1\}$};
  \node [] (l4_3) at (0,-4) {$\{q_1\}$};
  \node [] (l4_4) at (1,-4) {$\{q_1\}$};
  \node [] (l4_5) at (2,-4) {$\{q_0\}$};
  \node [] (w0) at (-3, -0.5) {$p$};
  \node [] (w1) at (-3, -1.5) {$\neg p$};
  \node [] (w2) at (-3, -2.5) {$p$};
  \node [] (w3) at (-3, -3.5) {$p$};
  \node [] (w4) at (-3, -4.5) {$\vdots$};
  \path[-]
    (root) edge (l1_1) edge (l1_2)
    (l1_2) edge (l2_2) edge (l2_3)
    (l2_2) edge (l3_2)
    (l2_3) edge (l3_3) edge (l3_4)
    (l3_2) edge (l4_2)
    (l3_3) edge (l4_3)
    (l3_4) edge (l4_4) edge (l4_5)
  ;
\end{tikzpicture}
}
\quad
\subfloat[][the reduced split tree of the NBW in Figure~\ref{fig:nbw} on the
accepted word $p \neg pp^\omega$]{\label{fig:reduced_split_tree}
\begin{tikzpicture}[
  auto, initial text=,
  semithick, >=stealth,
]
  \node [] (root) at (0,0) {$\{q_0\}$};
  \node [] (l1_1) at (-0.5,-1) {$\{q_1\}$};
  \node [] (l1_2) at (0.5,-1) {$\{q_0\}$};
  \node [] (l2_1) at (0,-2) {$\{q_1\}$};
  \node [] (l2_3) at (1,-2) {$\{q_0\}$};
  \node [] (l3_1) at (-0.5,-3) {$\{q_1\}$};
  \node [] (l3_4) at (1.5,-3) {$\{q_0\}$};
  \node [] (l4_1) at (-1,-4) {$\{q_1\}$};
  \node [] (l4_5) at (2,-4) {$\{q_0\}$};
  \node [] (w0) at (-3, -0.5) {$p$};
  \node [] (w1) at (-3, -1.5) {$\neg p$};
  \node [] (w2) at (-3, -2.5) {$p$};
  \node [] (w3) at (-3, -3.5) {$p$};
  \node [] (w4) at (-3, -4.5) {$\vdots$};
  \path[-]
    (root) edge (l1_1) edge (l1_2)
    (l1_2) edge (l2_1)
    (l1_2) edge (l2_3)
    (l2_1) edge (l3_1)
    (l2_3) edge (l3_4)
    (l3_1) edge (l4_1)
    (l3_4) edge (l4_5)
  ;
\end{tikzpicture}
}
\caption{Examples of a split tree and a reduced split tree\label{fig:trees}}
\end{figure}
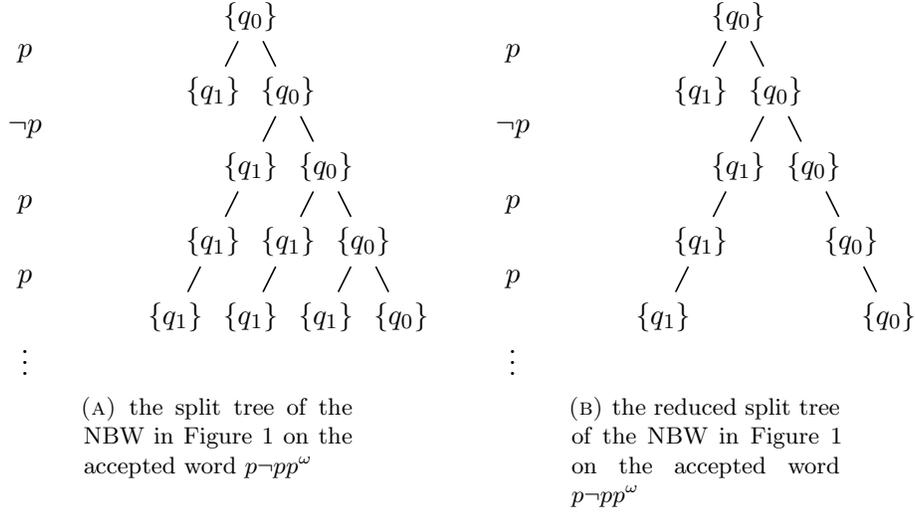

Consider the NBW in Figure~\ref{fig:nbw} where the alphabet is $\{p,
\neg p\}$, the initial state is $q_0$, and the acceptance condition is
$\{q_1\}$. 
The split tree and the reduced split tree of the NBW on the accepted
word $p \neg pp^\omega$ are shown in Figure~\ref{fig:trees}.

\section{Approaches to Complementation}\label{sec:review}

We review the four approaches to the complementation of \buchi
automata in this section.
In each approach, we identify one construction with the best
worst-case complexity.
The four identified constructions will be taken into account later in
our comparative experimentation.

\subsection*{Ramsey-based approach.}
The very first complementation construction introduced by \buchi in
1960 involves a Ramsey-based combinatorial argument and results in a
$2^{2^{O(n)}}$ blow-up in the state size~\cite{buchi:decision}.
This construction was later improved by Sistla, Vardi, and Wolper to reach 
a single-exponential complexity $2^{O(n^2)}$~\cite{sistla:complementation}.
The improved construction, referred to as \ramsey in this paper,
constructs a complement as the union of several NBWs.
Each NBW accepts a subset of the complement language in the form of
$UV^\omega$ where $U$ and $V$ are recognized respectively by two
classic finite automata on finite words.

Various optimization heuristics for the Ramsey-based approach were
described in \cite{abdulla:when,fogarty:efficient}, but the focus in these
works was on universality and containment.
In spite of the quadratic exponent of the Ramsey-based approach, it was shown in
\cite{abdulla:when,fogarty:buchi,fogarty:efficient} to be quite competitive
for universality and containment testing.
\hide{
The bottleneck of this approach is in
  constructing  a set of finite automata with infinite words with
  $4^{O(n^2)}$ states to answer two questions for every word $w$ and
  every two states $s$ and $t$ of the \buchi automaton. 
\begin{enumerate}
  \item Is there a run on $w$ starting with $s$ and ending with $t$?
  \item Is there a run on $w$ starting with $s$, ending with $t$, and
  passing some accepting state?
\end{enumerate} 
} 

\subsection*{Determinization-based approach.} Safra's $2^{O(n \log n)}$
construction is the first complementation construction that matches
the $\Omega(n!)$ lower bound~\cite{safra:complexity}.
The main idea is the use of (1) Safra trees to capture the history of
all runs on a word and  (2) marks to indicate whether a run passes an
accepting state again or dies.
Later on, Muller and Schupp introduced a similar determinization 
construction which records more information and yields larger
complements in most cases, but can be understood more
easily~\cite{muller:simulating,althoff:observations}.

The determinization-based approach performs complementation in stages:
first convert an NBW to a deterministic automaton, then complement
the deterministic automaton by modifying only the acceptance
condition, and finally convert the complement deterministic automaton
to an NBW.
Both Safra's construction and the construction by Muller and Schupp
use DRWs as the intermediate deterministic automata.
In~\cite{piterman:buchi}, Piterman improved Safra's construction by 
using a more compact tree structure and using DPWs as the intermediate
deterministic automata.
The improved construction by Piterman, referred to as
\piterman (or simply \SP) in this paper, yields an upper bound of
$n^{2n}$. (See also \cite{Sch09}.)

As the determinization-based approach performs complementation in
stages, different optimization techniques can be applied separately to
the different stages.
For instance, several optimization heuristics on
Safra's determinization and on simplifying the intermediate DRWs were
proposed by Klein and Baier~\cite{klein:experiments}.

\subsection*{Rank-based approach.} The rank-based approach, proposed
by Kupferman and Vardi, uses rank functions to measure the
progress made by a node of a run tree towards
fair termination~\cite{kupferman:weak}.
The basic idea of this approach may be traced back to Klarlund's
construction~\cite{klarlund:progress}.
Both constructions have complexity $2^{O(n \log n)}$.
There were also several optimization techniques proposed
in~\cite{gurumurthy:complementing,friedgut:buchi,karmarkar:minimal}.
A final improvement was proposed recently by
Schewe~\cite{schewe:buchi} to the construction
in~\cite{friedgut:buchi}.
The construction with Schewe's improvement, refered to as \rank in
this paper, performs a subset construction in the first phase.
From some point, it guesses ranks and transits from the first phase to
the second phase, where the guesses are verified.
Schewe proposed doing this verification in a piecemeal fashion.
This yields a complement with $O((0.76n)^n)$ states, which matches the 
known lower bound modulo an $O(n^2)$ factor.

Unlike the determinization-based approach that collects information from the
history, the rank-based approach guesses ranks bounded by $2(n -|\f|)$ and 
results in many nondeterministic choices.  This nondeterminism means
that the rank-based construction often creates more useless states because 
many guesses may be verified later to be incorrect.

\subsection*{Slice-based approach.} The slice-based construction was
proposed by K\"ahler and Wilke in 2008~\cite{kahler:complementation}.
The blow-up of the construction is $4(3n)^n$ while its preliminary
version in~\cite{vardi:automata07}, referred to as \slice here,
has a $(3n)^n$ blow-up\footnote{The construction
  in~\cite{kahler:complementation} has a higher complexity than its
  preliminary version because it aims at treating complementation and
  disambiguation in a uniform way.}. 
Unlike the determinization-based and the rank-based approaches that
analyze run trees, the slice-based approach analyzes reduced split
trees.
The construction \slice uses slices as states of the complement and
performs a construction based on the evolution of reduced split trees
in the first phase.
By decorating nodes in slices at some point, it guesses whether a node
belongs to an infinite branch of a reduced split tree or the node has
a finite number of descendants.
In the second phase, it verifies the guesses and enforces that
accepting states will not occur infinitely often.

The first phase of \slice in general creates more states than the first 
phase of \rank because of an ordering of nodes in the reduced split 
trees.
Similar to \rank, \slice also introduces nondeterministic choices 
in guessing the decorations.
While \rank guesses ranks bounded by $2(n - |\f|)$, \slice guesses
only the decorations from a fixed set of size 3.

\section{Comparison of Complementation Approaches}\label{sec:comparison}

Based on preliminary experiments~\cite{tsay:buchi_sttt}, we chose four
representative complementation constructions, namely 
\ramsey \cite{sistla:complementation},
\piterman~\cite{piterman:buchi}, \rank~\cite{schewe:buchi}, and
\slice~\cite{vardi:automata07}, that we considered the most
efficient in each approach.
These constructions were implemented in GOAL\footnote{We use the first
  generation of GOAL to perform all the
  experiments.}~\cite{tsay:goal,tsai:goal}.
In the following, we first describe our implementations, the settings
of experiments, and then present the experimental results.

\subsection{Implementations}

All the four implemented constructions use the same automaton
structure in GOAL.
Unlike modern model checkers that encode transition relations of
automata implicitly in BDD~\cite{bryant:graph}, GOAL represents
automata explicitly, that is, every transition is implemented as a
Java object.
The implemented constructions also use the same functions to access
the alphabet, the states, the transitions, and the acceptance
condition of an automaton.
An automaton in GOAL is a Java object with a HasSet of atomic
propositions (or classical symbols), a HashSet of states, an
initial state, a HashSet of transitions, and an acceptance condition.
A state in an automaton has a unique ID and a possibly empty label.
A transition in an automaton has a unique ID, a reference to the
source state of the transition, a reference to the destination state
of the transition, and a label.
A \buchi condition is a Vector of states.
A parity condition is a Vector of Vectors of states where a state in
the $i$-th vector has a parity $i$.
In order to access the successors and predecessors of a state quickly,
an automaton contains three HashMaps, namely \textit{fmap}, \textit{tmap},
and \textit{ftmap}, of which fmap maps a state to its outgoing
transitions, tmap maps a state to its incoming transitions, and ftmap
maps a state $s$ and a state $t$ to the transitions from $s$ to $t$.

Our implementations of complementation constructions basically
construct a complement incrementally from the initial state following
the description of the original
papers~\cite{sistla:complementation,piterman:buchi,schewe:buchi,vardi:automata07}.
As a state object in GOAL does not always match the state structure of
the complement in theory, we define additional Java classes to model
the state structure of the complement.
During a construction, two HashMaps are maintained respectively to map a
state object created in the complement to the underlying state
structure described in the paper and vice versa.
Our implementations do not use fancy data structures to represent the
underlying state structures of the complements.
For example, in \piterman, the state structure of the complements is
represented by a tree where a tree node contains an ArrayList of
states, a reference to its parent, references to its children, and
references to its older siblings.
In \rank, the state structure is represented by (1) TreeSets of states
(for the subset construction) and (2) tuples containing two TreeSets of
states, a ranking function represented as a HashMap object, and
an integer for the turn-wise cut-point
optimization~\cite{schewe:buchi}.

More details of our implementations can be obtained directly from the
source code, which is released with the first generation of
GOAL\footnote{The source code of the first generation of GOAL is
  released per request. Please visit http://goal.im.ntu.edu.tw/ for
  more details.}.

\subsection{Settings of Experiments}

We randomly generated 11,000 automata\footnote{A specialized version
  of GOAL and all the generated automata used in the experiments can
  be directly downloaded from http://goal.im.ntu.edu.tw/ without
  registration.} with an alphabet of size 2 and state sets of size 15
based on the approach proposed by Tabakov and
Vardi~\cite{tabakov:model}.
Among the 11,000 automata of state size 15, denoted by $\set{A}_{15}$,
each 100 automata were generated from a combination of 11 transition
densities (from 1.0 to 3.0) and 10 acceptance densities (from 0.1 to
1.0).
For every generated automaton $A = (\Sigma, Q, q_0, \delta, \f)$ with
$n$ states, symbol $a \in \Sigma$, transition density $r$, and
acceptance density $f$, we made $q \in \delta(p, a)$ for $\lceil rn
\rceil$ pairs of states $(p, q) \in Q^2$ uniformly chosen at random
and added $\lceil fn \rceil$ states to $\f$ uniformly at random.
Our parameters were chosen to generate a large set of complementation
problems, ranging from easy to hard.
The experiment was performed on a cluster at Rice University
(http://rcsg.rice.edu/sugar/int/).
For each complementation task, we allocated one 2.83-GHz CPU and 1 GB of
memory.
The timeout of a complementation task was 10 minutes.

\subsection{Experimental Results}

\begin{table}
\caption{A comparison of the four representative
  constructions based on $\set{A}_{15}$\label{exp:orig_size15} without
  and with preminimization.
  The preminimization is denoted by \texttt{+P}.
  We will use \SP as a shorthand of \piterman in all tables.}
\small
\begin{center}
\begin{tabular}{|r|r|r|c|rr|rr|r|}
\hline \multicolumn{1}{|c|}{Constructions} & \multicolumn{1}{|c|}{$T$} &
\multicolumn{1}{|c|}{$M$} & Eff. Samples & $S_{R}$ & (Win) & $S_{L}$ & (Win) &
\multicolumn{1}{|c|}{$S_L/S_R$} \\ \hline\hline
\multicolumn{9}{|l|}{$\set{A}_{15}$ (without preminimization)} \\ \hline
\ramsey   & 4,564 &    36 & 2,259 & 513.85 &     (0) & 30.82 & (522.50) & 0.060 \\
\SP &     5 &     0 &       &  45.26 & (1,843) &  2.27 & (556.67) & 0.050 \\
\rank     & 5,303 &     0 &       & 260.41 &   (415) &  2.79 & (649.17) & 0.011 \\
\slice    & 3,131 & 3,213 &       & 790.92 &     (1) &  3.03 & (530.67) & 0.004 \\ \hline
\hline
\multicolumn{9}{|l|}{$\set{A}_{15}$ (with preminimization)} \\ \hline
\ramseyp   & 4,190 &    41 & 3,522 & 193.72 &   (247) & 26.82 &   (776.25) & 0.218 \\
\SPp       &     4 &     0 &       &  11.08 & (1,712) &  2.31 &   (817.42) & 0.572 \\
\rankp     & 4,316 &     0 &       &  56.60 & (1,546) &  2.37 & (1,126.42) & 0.160 \\
\slicep    & 2,908 & 2,435 &       & 199.52 &    (17) &  2.94 &   (801.92) & 0.092 \\ \hline
\end{tabular}
\end{center}
\end{table}

The experimental results are summarized on the top of
Table~\ref{exp:orig_size15} where $T$ is the total number of timed-out
tasks and $M$ is the total number of tasks that run out of memory.
Compared to \piterman that has only 5 unfinished tasks, each of
\ramsey\footnote{\ramsey could not finish all tasks in our previous
  experiments in~\cite{tsai:state} because its implementation
  constructs the full state space. The implementation has been
  modified to construct only reachable states.}, \rank, and \slice has
around 50\% of unfinished tasks.
Besides the number of unfinished tasks, we also want to compare the
sizes of states of the constructed complements.
As these constructions may successfully finish different tasks, a 
construction may be misleadingly considered to be worse in producing
more states because it can finish harder tasks that have much larger
complements. 
Therefore, we only collect state-size information from 2,259
\emph{effective samples}, which are tasks finished successfully by all
the four constructions. 
Among the 2,259 effective samples, around 90\% of the automata are
universal.

The state-size information is shown in the columns $S_L$ and $S_R$.
The column $S_R$ is the average number of reachable states, while
$S_L$ is the average number of live states, of the complements.
The two columns show that \piterman is the best in average state
size.
The low $S_L/S_R$ ratio shows that \rank and \slice create more dead 
states that can be easily pruned off.

In addition to the number of unfinished tasks and the state-size
information, the number of smallest complements produced by a
construction among the effective samples is another measure of
performance.
A construction \emph{wins} in an effective sample w.r.t. $S_R$ (resp.,
$S_L$) if the complement produced by the construction is the smallest
in terms of reachable states (resp., live states).
The Win column of a construction in $S_R$ (resp., $S_L$) is the
fractional share of effective samples where the construction wins
w.r.t. $S_R$ (resp., $S_L$). 
If $k$ constructions win in an effective sample, each gets $1/k$
shares. 
The Win columns show that although \rank generates more dead states,
it produces more complements that are the smallest after pruning dead
states.

\begin{table}[htb]
\caption{A comparison of the four representative
  constructions based on the nonuniversal automata in
  $\set{A}_{15}$\label{exp:orig_nonuniv_15}}
\small
\begin{center}
\begin{tabular}{|r|c|rr|rr|r|}
\hline \multicolumn{1}{|c|}{Constructions} & Eff. Samples & S$_{R}$ &
(Win) & S$_{L}$ & (Win) & \multicolumn{1}{|c|}{S$_L$/S$_R$} \\
\hline\hline
\multicolumn{7}{|l|}{$\set{A}_{15}$ (without preminimization)} \\ \hline
\ramsey   & 171 & 1,892.37 &     (0) & 397.10 &      (0) & 0.210 \\
\SP       &     &    36.38 & (102.5) &  17.77 &  (34.67) & 0.488 \\
\rank     &     &   156.63 &  (67.5) &  24.61 & (127.67) & 0.157 \\
\slice    &     &   422.88 &     (1) &  27.75 &   (8.67) & 0.066 \\ \hline
\hline
\multicolumn{7}{|l|}{$\set{A}_{15}$ (with preminimization)} \\ \hline
\ramseyp   & 418 & 1,000.89 &   (0.0) & 218.55 &   (0.25) & 0.218 \\
\SPp       &     &    21.02 & (116.0) &  12.02 &  (41.42) & 0.572 \\
\rankp     &     &    77.97 & (300.5) &  12.51 & (350.42) & 0.160 \\
\slicep    &     &   189.90 &   (1.5) &  17.39 &  (25.92) & 0.092 \\ \hline
\end{tabular}
\end{center}
\end{table}

\rank and \slice become much closer to \piterman in $S_L$ because
around 90\% of the 2,259 effective samples are universal automata,
whose complements have no live states. 
If we only consider nonuniversal automata, the gaps between \piterman
and \rank, and \piterman and \slice in $S_L$ become larger as
shown on the top of Table~\ref{exp:orig_nonuniv_15}.

As the heuristic of preminimization applied to the input automata,
denoted by \texttt{+P}, is considered to help the nondeterministic
constructions more than the deterministic one, we also
compare the four constructions with preminimization and summarize the
results on the bottom of Tables~\ref{exp:orig_size15}
and~\ref{exp:orig_nonuniv_15}.
We only applied the preminimization implemented in GOAL
tool, namely the simplification by simulation
in~\cite{somenzi:efficient}.
According to our experimental results, the preminimization does
improve \ramsey, \rank, and \slice more than \piterman in the
complementation but does not close too much the gap between them in
the comparison, though there are other preminimization techniques that
we did not apply in the experiment.

In summary, the experimental results show that (1) \piterman is the
best in average state size and in the number of finished tasks, (2)
\ramsey is not competitive in complementation even though it is
competitive in universality and containment testing as shown
in~\cite{abdulla:when,fogarty:buchi,fogarty:efficient}, and (3) \slice
has the most unfinished tasks (even more than \ramsey) and, compared
to \piterman and \rank, produces many more states.
As we will show later, we can improve the performance of \slice
significantly by employing various optimization heuristics.

Besides the 11,000 automata with 15 states, another 11,000 automata
with 10 states and another 11,000 automata with 20 states, denoted by
$\set{A}_{10}$ and $\set{A}_{20}$ respectively, were also used in our
experiments.
We include the experimental results based on $\set{A}_{10}$ and
$\set{A}_{20}$ in the appendix because $\set{A}_{10}$ contains fewer
tough automata especially for \piterman while $\set{A}_{20}$ is
too hard to get effective samples.
Moreover, the comparisons based on $\set{A}_{10}$ and $\set{A}_{20}$
are basically consistent to those based on $\set{A}_{15}$.

\section{Optimization Techniques}\label{sec:optimization}

The following optimization heuristics are described in this section:
simplifying DPWs by simulation (\texttt{+S}) and merging equivalent
states (\texttt{+E}) for \piterman; maximizing the \buchi acceptance
set (\texttt{+A}) for \rank; deterministic decoration (\texttt{+D}),
reducing transitions (\texttt{+R}), and merging adjacent nodes
(\texttt{+M}) for \slice.
For \piterman , the first heuristic \texttt{+S} is applied to an
intermediate complement DPW and yields an NPW while the second
heuristic \texttt{+E} is applied to the conversion from an NPW to an
NBW.
The heuristic \texttt{+A} for \rank is applied to the input NBW before
complementation and may be also usefull for other constructions.
The three heuristics \texttt{+D}, \texttt{+R}, and \texttt{+M} for
\slice are applied to the complementation construction.

\subsection{For \piterman}\label{sec:opt-piterman}

\piterman performs complementation via several intermediate stages:
starting with the given NBW, it computes first an equivalent DPW, then
a complement DPW, and finally a complement NBW.
We address (1) the simplification of the complement DPW, which results
in an NPW, and (2) the conversion from an NPW to an equivalent NBW.

\paragraph{\emph{Simplifying DPWs by simulation} (\texttt{+S})}
For the simplification of the complement DPW, we borrow from the ideas
of Somenzi and Bloem~\cite{somenzi:efficient}.
The direct and reverse simulation relations they introduced are useful
in removing transitions and possibly states of an NBW while retaining
its language. 
We define the simulation relations for an NPW in order to apply the
same simplification technique.
Let $(\Sigma, Q, q_0, \delta, \f)$ be an NPW.
Define a predecessor function $\delta^{-1}(p, a) = \{q \mid p \in
\delta(q, a)\}$ for $p \in Q$ and $a \in \Sigma$.
Given $p, q \in Q$, $p$ is \textit{directly simulated} by $q$ iff (1)
for all $p' \in \delta(p, a)$, there is $q' \in \delta(q, a)$ such
that $p'$ is directly simulated by $q'$, and (2) $\f(p) = \f(q)$.
Similarly, $p$ is \textit{reversely simulated} by $q$ iff (1) for all
$p' \in \delta^{-1}(p, a)$, there is $q' \in \delta^{-1}(q,
a)$ such that $p'$ is reversely simulated by $q'$, (2) $\f(p) =
\f(q)$, and (3) $p$ = $q_0$ implies $q$ = $q_0$. 
After simplification using simulation relations, as
in~\cite{somenzi:efficient}, a DPW may become nondeterministic.
Since the complementation of a DPW is much easier than that of an
NPW, the simplification by simulation is applied to the complement DPW
(in the second stage) but not to the equivalent DPW (in the first
stage).

\paragraph{\emph{Merging equivalent states (\texttt{+E})}}
As for the conversion from an NPW to an NBW, a typical way in the
literature is to direcly apply the conversion from an NRW to an 
NBW~\cite{king:complexity,gradel:automata} because the parity
condition is a special case of the Rabin condition.
Here we first review the conversion from an NRW to an NBW adapted for
an NPW.

Intuitively, the conversion nondeterministically guesses the minimal even parity
passed infinitely often in a run starting from some state. 
Once a run is guessed to pass a minimal even parity $2k$ infinitely
often starting from a state $p$, every state $q$ in the run after $p$ should
have a parity greater than or equal to $2k$ and $q$ is designated as
an accepting state in the resulting NBW if it has parity $2k$.

Given an NPW $P = (\Sigma, Q, q_0, \delta, \f)$ where $\f : Q
\rightarrow \{0, 1, \ldots, 2r\}$, the typical conversion constructs
the equivalent NBW $A = (\Sigma, S, s_0, \Delta, \g)$ where
\begin{itemize}
  \item $S = Q \times \{0, 2, \ldots, 2r\}$,
  \item $s_0 = (q_0, 0)$,
  \item $\Delta : S \times \Sigma \rightarrow 2^S$ is the transition function
    satisfying the following two conditions:
  \begin{itemize}
    \item $(q_j, 2k) \in \Delta((q_i, 0), a)$ iff $k > 0$ and $q_j \in
      \delta(q_i, a)$
    \item $(q_j, 2k) \in \Delta((q_i, 2k), a)$ iff $q_j \in
      \delta(q_i, a)$ and $\f(q_j) \geq 2k$, and
  \end{itemize}
  \item $\g = \{(q, 2k) \in S \mid \f(q) = 2k \}$.
\end{itemize} 
A run of $A$ will always look like $(q_0,
0)\cdots(q_{i-1},0)(q_i,2k)(q_{i+1},2k)\cdots$ where the transitions
from $(q_{i-1},0)$ to $(q_i,2k)$ represent the guess of $2k$ to be the
minimal even parity passed infinitely often and from there on $2k$
remains unchanged.

\begin{lem}
Given an NPW $P$, the typical conversion constructs an NBW $A$ such that
$L(P) = L(A)$.
\end{lem}
\proof
The typical conversion basically follows the conversion in
\cite{king:complexity} but restricts the Rabin condition to a Rabin
chain, which is equivalent to the parity condition of $P$.
Thus, the proof in \cite{king:complexity} applies. \qed

We propose to perform the conversion with states merged and with the start of
guessing the minimal even parity $2k$ delayed to a state that has the even
parity.
Let $P = (\Sigma, Q, q_0, \delta, \f)$ be an NPW where $\f : Q
\rightarrow \{0, 1, \ldots, 2r\}$.
We first define an equivalence relation on states with respect to
an even parity in order to merge the states in the conversion.
Two states $p$ and $q$ are equivalent with respect to an even
parity $2k$, denoted by $p \equiv_{2k} q$, iff $\delta(p, a) =
\delta(q, a)$ for all $a \in \Sigma$, and either 
\begin{itemize}
  \item $\f(p) = \f(q) = 2k$,
  \item $\f(p) > 2k$ and $\f(q) > 2k$, or
  \item $\f(p) < 2k$ and $\f(q) < 2k$.
\end{itemize}
Let $[q]_{2k} = \{ p \mid p \equiv_{2k} q \}$ be the equivalence class of
a state $q$ for a parity $2k$.
Let $[Q] = \{[q]_{2k} \mid q \in Q \mbox{ and } 0 \leq k \leq r \}$ denote
the set of equivalence classes of states in $Q$ for all even
parities.

Given an NPW $P = (\Sigma, Q, q_0, \delta, \f)$ where $\f : Q
\rightarrow \{0, 1, \ldots, 2r\}$, the improved conversion constructs
the equivalent NBW $A' = (\Sigma, S, s_0, \Delta, \g)$ where
\begin{itemize}
  \item $S = [Q] \times \{0, 2, \ldots, 2r\}$,
  \item $s_0 = ([q_0]_0, 0)$,
  \item $\Delta : S \times \Sigma \rightarrow 2^S$ is the transition
    function where
  \begin{description}
    \item[{[\rulename{TR1}]}] $([q]_{2k}, 2k) \in \Delta(([p]_0, 0),
      a)$ iff $k > 0$, $\delta(p, a) \cap [q]_{2k} \neq \emptyset$,
      and $\f(q) = 2k$,
    \item[{[\rulename{TR2}]}] $([q]_{2k}, 2k) \in \Delta(([p]_{2k},
      2k), a)$ iff $\delta(p, a) \cap [q]_{2k} \neq \emptyset$ and
      $\f(q) \geq 2k$, and
  \end{description}
  \item $\g = \{ ([q]_{2k}, 2k) \in S \mid \f(q) = 2k \}$.
\end{itemize}

\begin{lem}
\label{lemma:eq1}
If a word $w$ is accepted by $P$, then it is accepted by $A'$.
\end{lem}
\proof
Let $w$ be a word accepted by $P$ and $\rho$ an accepting run $q_0
q_1 \cdots$ of $P$ on $w$.
Suppose $2k$ is the minimal even parity passed infinitely often in
$\rho$ after some state $q_i$ of parity $2k$.
By the construction, there is a run $\rho' = [q_0]_0 [q_1]_0 \cdots
[q_{i-1}]_0 [q_i]_{2k} [q_{i+1}]_{2k} \cdots$ of $A'$ on $w$.
The transitions before and after $[q_i]_{2k}$ follow the transition rules
\rulename{TR1} and \rulename{TR2} respectively.
Since $2k$ is the minimal even parity passed infinitely often in
$\rho$, there are infinitely many $[q_m]_{2k}$'s ($i \leq m$) in
$\rho'$ such that $\f(q_m) = 2k$ and $[q_m]_{2k} \in \g$.
Thus, $\rho'$ is an accepting run of $A'$ on $w$.\qed

\begin{lem}
\label{lemma:eq2}
If a word $w$ is accepted by $A'$, then it is accepted by $P$.
\end{lem}
\proof
Let $w = a_0 a_1 \cdots$ be a word accepted by $A'$ and $\rho$ an
accepting run $[q_0]_0 [q_1]_0 \cdots [q_{i-1}]_0$\linebreak$[q_i]_{2k}
[q_{i+1}]_{2k} \cdots$ of $A'$ on $w$.
Let $M$ be an infinite set of indices such that $m \in M$ iff
$[q_m]_{2k} \in \g$.
By the construction, we can find a state $q_1' \in [q_1]_0$
such that $q_1' \in \delta(q_0, a_0)$.
Starting from $q_1'$, by the construction and by the definition of
equivalence classes, we can find a state $q_2' \in [q_2]_0$ such that
$q_2' \in \delta(q_1', a_1)$ and so on.
Therefore, there is a run $\rho' = q_0 q_1' q_2' \cdots q_i' q_{i+1}'
\cdots$ of $P$ on $w$ such that $q_j' \in [q_j]_0$ for $0 < j < i$ and
$q_j' \in [q_j]_{2k}$ for $j \geq i$.
By the transition function and the equivalence relation, $\f(q_j') \geq 2k$ for
$j \geq i$ and $\f(q_m') = 2k$ for all $m \in M$.
Hence, $2k$ is the minimal even parity passed infinitely often in
$\rho'$ after $q_i$ and $\rho'$ is an accepting run of $P$ on
$w$. \qed

\begin{thm}
$L(P) = L(A')$.
\end{thm}
\proof
The result follows directly from Lemmas \ref{lemma:eq1} and
\ref{lemma:eq2}. \qed
\newpage

\subsection{For \rank}\label{sec:rank}

\paragraph{\emph{Maximizing the \buchi acceptance set (\texttt{+A})}}.
As stated in Section~\ref{sec:review}, the ranks for the rank-based
approach are bounded by $2(n - |\f|)$.
The larger $\f$ is, the fewer the ranks are.
Thus, we propose to maximize the acceptance set of the input NBW before
complementation based on the following theorem without changing its
language, states, or transition function.

\begin{thm}
Let $A = (\Sigma, Q, q_0, \delta, \f)$ and $A' = (\Sigma, Q, q_0,
\delta, \g)$ be two NBWs where $\g \supseteq \f$.
Then, $L(A) = L(A')$ if for all $q \in \g$, every elementary cycle
containing $q$ also contains at least one state in $\f$.
\label{thm:maxacc}
\end{thm}
\proof
Since $\g \supseteq \f$, $L(A) \subseteq L(A')$.
To prove $L(A) \supseteq L(A')$, first let $\rho = q_0 q_1 \cdots$
be an accepting run of $A'$ on some word $w$.
Since $\rho$ is accepting, there exist some state $q_i \in \rho$ and
infinite indices $i_0 < i_1 < i_2 < \cdots$ such that $q_i \in \g$ and
$q_i = q_{i_0} = q_{i_1} = q_{i_2} = \cdots$.
For all $j \geq 0$, the sequence of states $q_{i_j} q_{i_j+1} \ldots
q_{i_{j+1}}$ forms a cycle, denoted by $C_j$.
As $q_i \in \g$ and a cycle is formed by elementary cycles, in each
$C_j$, there is some state $q_{k_j} \in \f$.
Since there are infinitely many $C_j$'s but $Q$ is finite, there
exists some state $q_k \in \f$ occurring infinitely many times in
$\rho$.
Thus, $\rho$ is an accepting of $A$ on $w$ and $L(A) \supseteq
L(A')$. \qed

The elementary cycles of an automaton can be found by the algorithm
in~\cite{johnson:finding} with a time complexity $O((n + e)(c +
1))$ and a space complexity $O(n + e)$ where $n$ is the number of
states, $e$ the number of transitions, and $c$ the number of
elementary cycles in the automaton\footnote{Instead of finding
  elementary cycles, our implementation makes a state accepting if the
  state cannot go back to itself without passing an accepting state in
  $\f$, which is checked by a depth-first search.}.

This heuristic can also be applied to other complementation
approaches as it maximizes the acceptance set of the input NBW before
complementation.
We will show the improvement made by this heuristic for \piterman,
\rank, and \slice later in Section~\ref{sec:experiment}.

\subsection{For \slice}\label{sec:opt-slice}

The central idea of \slice is based on the following
lemma~\cite{kahler:complementation}.
\begin{lem}
\label{lemma:slice-cutoff}
A word $w$ is rejected by an NBW $A$ iff the reduced split tree of $A$
on $w$ has a \emph{cutoff}, which is a level $i$ such that after the
$i$-th level in the reduced split tree, all the left children are in
finite branches.
\end{lem}
Note that a reduced split tree may have infinitely many cutoffs.

\begin{figure}
\begin{center}
\begin{tikzpicture}[
  auto, initial text=,
  semithick, >=stealth,
]
  \node [] (root) at (0,0) {$\{q_0\}$};
  \node [] (l1_1) at (-.5,-1) {$\{q_1\}^0$};
  \node [] (l1_2) at (.5,-1) {$\{q_0\}^1$};
  \node [] (l2_1) at (0,-2) {$\{q_1\}^*$};
  \node [] (l2_2) at (1,-2) {$\{q_0\}^1$};
  \node [] (l3_1) at (-.5,-3) {$\{q_1\}^0$};
  \node [] (l3_2) at (1.5,-3) {$\{q_0\}^1$};
  \node [] (l4_1) at (1,-4) {$\{q_1\}^*$};
  \node [] (l4_2) at (2,-4) {$\{q_0\}^1$};
  \node [] (w0) at (-1.5, -0.5) {$p$};
  \node [] (w1) at (-1.5, -1.5) {$\neg p$};
  \node [] (w2) at (-1.5, -2.5) {$p$};
  \node [] (w3) at (-1.5, -3.5) {$\neg p$};
  \node [] (w4) at (-1.5, -4.5) {$\vdots$};
  \path[-]
    (root) edge (l1_1) edge (l1_2)
    (l1_2) edge (l2_1) edge (l2_2)
    (l2_1) edge (l3_1)
    (l2_2) edge (l3_2)
    (l3_2) edge (l4_1) edge (l4_2)
  ;
\end{tikzpicture}
\end{center}
\caption{A decorated reduced split tree of the NBW in Figure~\ref{fig:nbw} on
  the rejected word $(p \neg p)^\omega$\label{fig:rejected_rst}}
\end{figure}
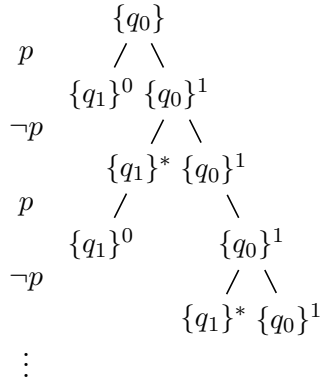

As an example, the reduced split tree of the NBW in
Figure~\ref{fig:nbw} on a rejected word $(p \neg p)^\omega$ is shown
in Figure~\ref{fig:rejected_rst} where the superscripts $0$, $*$, and
$1$ are decorations to explained later.
It can be seen that after the first level of the reduced split tree,
all the accepting states of the NBW are in finite branches.

Based on Lemma~\ref{lemma:slice-cutoff}, \slice constructs a complement with
slices as states to accept all reduced split trees of an input NBW on rejected
words by guessing the cutoffs nondeterministically.
A decoration scheme is applied to verify whether a slice is on a cutoff.
The transition relation of the complement is divided into two phases.
In the first phase, the transition relation is based on the evolution
of reduced split trees.
When the construction nondeterministically chooses a slice as the
slice on some cutoff, it guesses the decorations of the nodes in the
slice and goes from the first phase to the second phase, where the
decorations are verified.

\begin{figure}[htb]
\begin{center}
\begin{tabular}{rcl}
\vspace{1em}
not reset slice: & \quad\quad &
\subfloat[][\rulename{D1}]{\label{fig:basic_scheme_1}
\begin{tikzpicture}[
  auto, initial text=,
  semithick, >=stealth,
  must/.style={circle,draw},
  may/.style={circle,draw,dashed}
]
  \node[must] (q1) at (1,1) {$1$};
    \node[may] (q1s) at (0.5,0) {$*$};
    \node[must] (q11) at (1.5,0) {$1$};
  \path[->]
    (q1) edge (q1s) edge (q11)
  ;
\end{tikzpicture}
}
\quad
\subfloat[][\rulename{D2-1}]{\label{fig:basic_scheme_0}
\begin{tikzpicture}[
  auto, initial text=,
  semithick, >=stealth,
  must/.style={circle,draw},
  may/.style={circle,draw,dashed}
]
  \node[must] (q0) at (5,1) {$0$};
    \node[may] (q00) at (4.5,0) {$0$};
    \node[may] (q01) at (5.5,0) {$0$};
  \path[->]
    (q0) edge (q00) edge (q01)
  ;
\end{tikzpicture}
}
\quad
\subfloat[][\rulename{D2-2}]{\label{fig:basic_scheme_s}
\begin{tikzpicture}[
  auto, initial text=,
  semithick, >=stealth,
  must/.style={circle,draw},
  may/.style={circle,draw,dashed}
]
  \node[must] (qs) at (3,1) {$*$};
    \node[may] (qs0) at (2.5,0) {$*$};
    \node[may] (qs1) at (3.5,0) {$*$};
  \path[->]
    (qs) edge (qs0) edge (qs1)
  ;
\end{tikzpicture}
}
\\ \hline \\
reset slice: & \quad\quad &
\subfloat[][\rulename{D3}]{\label{fig:basic_scheme_1r}
\begin{tikzpicture}[
  auto, initial text=,
  semithick, >=stealth,
  must/.style={circle,draw},
  may/.style={circle,draw,dashed}
]
  \node[must] (q1r) at (9,1) {$1$};
    \node[may] (q1r0) at (8.5,0) {$0$};
    \node[must] (q1r1) at (9.5,0) {$1$};
  \path[->]
    (q1r) edge (q1r0) edge (q1r1)
  ;
\end{tikzpicture}
}
\quad
\subfloat[][\rulename{D4}]{\label{fig:basic_scheme_sr}
\begin{tikzpicture}[
  auto, initial text=,
  semithick, >=stealth,
  must/.style={circle,draw},
  may/.style={circle,draw,dashed}
]
  \node[must] (qsr) at (7,1) {$*$};
    \node[may] (qsr0) at (6.5,0) {$0$};
    \node[may] (qsr1) at (7.5,0) {$0$};
  \path[->]
    (qsr) edge (qsr0) edge (qsr1)
  ;
\end{tikzpicture}
}
\end{tabular}
\end{center}

\caption{The decoration rules \rulename{D1}, \rulename{D2}
  (which consists of \rulename{D2-1} and \rulename{D2-2}),
  \rulename{D3}, and \rulename{D4} applied in the basic \slice
  construction.
  The upper three rules are applied only to non-reset slices while the
  lower two rules only to reset slices.
  A node in a slice is represented by a circle in which the
  label denotes the decoration of the node.
  As an example, when rule \rulename{D3} is applied to a reset slice,
  the left child of an $1$-node is decorated by $0$ while the right
  child is decorated by $1$.
  A child can be absent when applying a rule if it is dashed and
  otherwise it must exist.
  \label{fig:basic_scheme}
}
\end{figure}
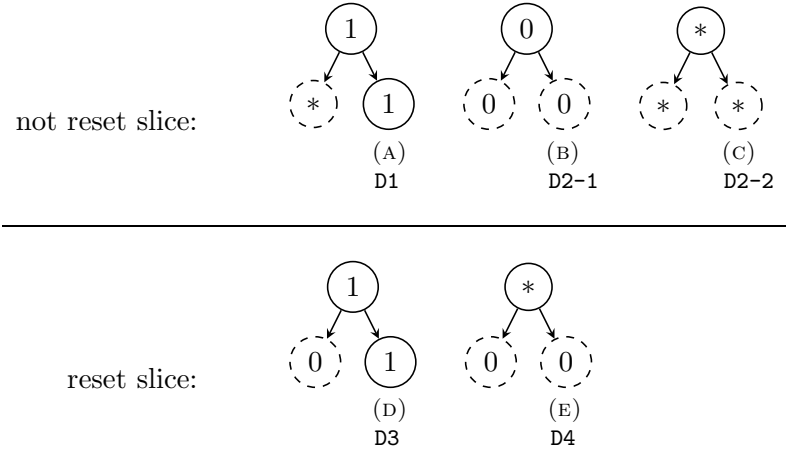

A node in a slice can be decorated by $1$, $0$, or $*$.
Intuitively, the decoration $1$ indicates that a node is in an
infinite branch of a reduced split tree.
The decoration $0$ indicates that the descendants of a node die out
eventually before the next \emph{reset slice}, which is a slice with
no node decorated by $0$.
The decoration $*$ has the same meaning as $0$ but the check is put on
hold after the next reset slice where the children of $*$-nodes are
decorated by $0$.
Shown in Figure~\ref{fig:basic_scheme} are the decoration rules, which
enforce that when reset slices are passed infinitely many times, descendants
of the left children after decoration will eventually die out.

The reason why both $0$ and $*$ are used for nodes on finite
branches is that when we focus on the sequence of slices that form a
reduced split tree, we want to distinguish a node that just died on
the previous level between the same node that is just born on the
current level.
Otherwise, reset slices may not appear if nodes decorated by $0$ are
always born on a level immediately after they died out on the previous
level.
For example, consider the decorated reduced split tree in
Figure~\ref{fig:rejected_rst}.
If the decoration $*$ is replaced by $0$, the reduced split tree will
no longer contain reset slices because $\{q_1\}^0$ appears on every
level after the root, but actually $\{q_1\}^0$ dies on every odd 
level and is born on every even level.


Before the formal description of the basic \slice construction, we first
introduce some notations.
Let $A = (\Sigma, Q, q_0, \delta, \f)$ be an NBW and $D$ the set $\{0, *, 1\}$.
An \emph{undecorated slice} over $Q$ is a finite, pairwise disjoint, 
sequence $Q_0 \cdots Q_{n-1}$ of non-empty subsets of $Q$. 
A \emph{decorated slice} over $Q$ is a finite
sequence $(Q_0, d_0)\cdots(Q_{n-1}, d_{n-1})$ where $Q_0 \cdots
Q_{n-1}$ form an undecorated slice and $d_i \in D$ for $i < n$.
The $i$-th node of a slice $s$ is denoted by $s(i)$ and the number of
nodes of a slice is denoted by $|s|$.
The empty sequence, denoted by $\bot$, is a special slice
considered both undecorated and decorated.
The set of slices over $Q$ is denoted by $S = S^u \cup S^d$ where
$S^u$ is the set of undecorated slices and $S^d$ the set of decorated
slices.
Let $s = (Q_0, d_0)\cdots(Q_{n-1}, d_{n-1}) \in S^d$.
Define $\projq{s} = Q_0 \cdots Q_{n-1}$ to be the undecorated version
of $s$ and $\projd{s} = \{d_0, \ldots, d_{n-1}\}$.
We say $s$ is a \emph{reset slice} iff $0 \not\in \projd{s}$
and $s$ is \emph{doomed} iff $1 \not\in \projd{s}$.
In particular, $\bot$ is a reset slice and is doomed.

\subsubsection{The basic \slice construction.}
\enlargethispage{\baselineskip}

Let $A = (\Sigma, Q, q_0, \delta, \f)$ be an NBW.
The complement constructed by \slice is $A' = (\Sigma, S, s_0, \Delta,
\g)$ where
\begin{itemize}
\item $s_0 = \{q_0\}$,
\item $\Delta = S \times \Sigma \rightarrow 2^S$ is the transition
  function described below, and
\item $\g = \{s \in S^d \mid s \mbox{ is a reset slice}\}$.
\end{itemize}
For all $s \in S$ and $a \in \Sigma$, $\Delta(s, a)$ is defined as
follows:
\begin{itemize}
\item $\Delta(s, a) = \{ \delta_u(s, a) \} \cup \delta_g(s, a)$ if $s \in
  S^u$.
\item $\Delta(s, a) = \{ \delta_d(s, a) \}$ if $s \in S^d$.
\end{itemize}
The functions $\delta_u$, $\delta_g$, and $\delta_d$ correspond
respectively to the transition functions in the first phase, from the
first phase to the second phase, and in the second phase.
\begin{itemize}
\item
The transition function $\delta_u : S^u \times \Sigma \rightarrow S^u$
represents the first phase of \slice with $\delta_u(s, a)$ giving the
next level of $s$ in a reduced split tree with respect to the symbol
$a$.
Let $s = Q_0 \cdots Q_{n-1} \in Q^u$ and $a \in \Sigma$.
Define $s' = Q_0' \cdots Q_{2n-1}'$ such that for $i < n$,
\begin{itemize}
\item $Q_{2i}' = (\cup_{q \in Q_i} \delta(q, a) \cap \f) - \bigcup_{j < 2i} Q_{j}'$, and 
\item $Q_{2i+1}' = (\cup_{q \in Q_i} \delta(q, a) - \f) - \bigcup_{j < 2i} Q_{j}'$.
\end{itemize}
By removing $\emptyset$ from $s'$, we can find $j_0 < \cdots <
j_{r-1}$ such that $\{j_0, \ldots, j_{r-1}\} = \{j < 2n \mid Q_j' \neq
\emptyset\}$.
The result $Q_{j_0} \cdots Q_{j_{r-1}}$ is called an \emph{$a$-successor} of
$s$, denoted by $\delta_u(s, a)$.
\item
The transition function $\delta_g : S^u \times \Sigma \rightarrow
2^{S^d}$ is applied when \slice nondeterministically goes from the first phase
to the second phase and starts decoration.
In this transition, it labels the children of an undecorated slice
nondeterministically by $0$ or $1$.
Thus for $s \in S^u$, $a \in \Sigma$, and $s' \in S^d$, $s' \in
\delta_g(s, a)$ iff $\projq{s'} = \delta_u(s, a)$ and 
$\projd{s'} \subseteq \{0, 1\}$.
\item
The transition function $\delta_d : S^d \times \Sigma \rightarrow S^d$
represents the second phase of \slice where it verifies the
decorations by evolving decorated slices in the following way.
Let $s = (Q_0, d_0) \cdots (Q_{n-1}, d_{n-1}) \in S^d$, $a \in
\Sigma$, and $s' = (Q_{j_0}', d_{j_0}') \cdots (Q_{j_r-1}',
d_{j_r-1}') \in S^d$ where $j$'s and $Q_j'$'s are defined as in the
definition of $\delta_u$, i.e., $\projq{s'} = \delta_u(\projq{s},
a)$.
The decorated slice $s'$ is an $a$-successor of $s$, denoted by
$\delta_d(s, a)$, iff the following two conditions are satisfied:
\begin{description}
  \item[{[\rulename{C1}]}] for all $i < n$ with $d_i = 1$, $Q_{2i+1}'
    \neq \emptyset$, 
  \item[{[\rulename{C2}]}] $d_j'$'s are decorated by the following
    rules: 
    \smallskip
    \begin{description}
    \item[{[\rulename{D1}]}] If $s$ is not a reset slice and $d_i =
      1$, then $d_{2i}' = *$ and $d_{2i+1}' = 1$.
    \item[{[\rulename{D2}]}] If $s$ is not a reset slice and $d_i \in
      \{0, *\}$, then $d_{2i}' = d_i$ and $d_{2i+1}' = d_i$.
    \item[{[\rulename{D3}]}] If $s$ is a reset slice and $d_i = 1$,
      then $d_{2i}' = 0$ and $d_{2i+1}' = 1$.
    \item[{[\rulename{D4}]}] If $s$ is a reset slice and $d_i = *$,
      then $d_{2i}' = 0$ and $d_{2i+1}' = 0$.
    \end{description}
\end{description}\medskip
\end{itemize}

\begin{thm}~\cite{vardi:automata07}
Given an NBW $A$, the basic \slice construction produces an
NBW $A'$ with $L(A') = \overline{L(A)}$.
\end{thm}

\subsubsection{The improved \slice construction.} 

We first describe three optimization heuristics applied to the basic \slice
construction and then the resulting improved \slice construction.

\paragraph{\emph{Deterministic decoration (\texttt{+D})}}
The first heuristic uses $1$ to label nodes that \emph{may} (rather
than \emph{must}) be in an infinite branch of a reduced split tree and
only verifies the condition \rulename{C2} in the second phase. 
Thus, all nodes could be decorated by $1$ in the guesses.
However, since the first evolution of the second phase always labels
a left (accepting) child by $0$ and a right (nonaccepting) child by
$1$, we actually decorate accepting nodes by $0$
and nonaccepting nodes by $1$ in the guesses.
Formally, let $A = (\Sigma, Q, q_0, \delta, \f)$ be an NBW.
We define $\delta_g'  : S^u \times \Sigma \rightarrow S^d$ and
$\delta_d'  : S^d \times \Sigma \rightarrow S^d$, which refine
respectively $\delta_g$ and $\delta_d$ based on this heuristic.
\begin{itemize}
\item
Let $s = Q_0 \cdots Q_{n-1} \in S^u$, $a \in \Sigma$, and $s' =
(Q_{j_0}', d_{j_0}') \cdots (Q_{j_{r-1}}', d_{j_{r-1}}') \in S^d$
where $j$'s and $Q_j'$'s are defined as in the basic \slice construction,
i.e., $\projq{s'} = \delta_u(s, a)$.
Then $s' = \delta_g'(s, a)$ iff for all $i < n$, $d_{2i}' = 0$ and $d_{2i+1}' =
1$.
\item
The transition function $\delta_d'$ is the same as $\delta_d$ in
the basic \slice construction except that the condition C1 of $\delta_d$
is not required to be satisfied.
\end{itemize}
This heuristic results in deterministic decoration.
The only nondeterminism comes from choosing when to start decorating.

\paragraph{\emph{Reducing transitions (\texttt{+R})}}
The second heuristic relies on the observation that if a run ends up
in the empty sequence $\bot$, the run will stay in $\bot$ forever and
we never need to decorate the run because it can reach $\bot$ without
any decoration. 
Thus we do not allow transitions from decorated slices other than
$\bot$ to $\bot$ or from any slice to doomed slices other than $\bot$;
recall that a slice is doomed if it has no node labeled by $1$,
i.e., every run through a doomed slice is expected to reach $\bot$.
\enlargethispage{\baselineskip}

\paragraph{\emph{Merging adjacent nodes (\texttt{+M})}}
The third heuristic recursively merges adjacent nodes decorated all by
$0$ or all by $*$.
The observation is that they are all guessed to have a finite number of
descendants and their successors will have the same decoration, either $0$ or
$*$.
Let $s = (Q_0, d_0)\cdots(Q_{n-1}, d_{n-1}) \in S^d$.
Based on this heuristic, we can recursively merge adjacent nodes $(Q_i, d_i)$
and $(Q_{i+1}, d_{i+1})$ in $s$ when $d_i = d_{i+1} = 0$ or $d_i = d_{i+1} = *$.
We call the result a \textit{merged slice} of $s$ and denote it by $\merge(s)$.

Given an NBW $A = (\Sigma, Q, q_0, \delta, \f)$ , the improved \slice
with all optimization heuristics in this section constructs the complement $A' =
(\Sigma, S, s_0, \Delta, \g)$ where
\begin{itemize}
\item $s_0 = \{q_0\}$,
\item $\Delta = S \times \Sigma \rightarrow 2^S$ is the transition
  function described below, and
\item $\g = \{s \in S^d \mid s \mbox{ is a reset slice}\}$.
\end{itemize}
For all $s \in S$ and $a \in \Sigma$, $\Delta(s, a)$ is defined as
follows:
\begin{itemize}
\item $\Delta(s, a) = \{ \delta_u(s, a) \}$ if $s \in S^u$ and
  $\delta_g'(s, a)$ is doomed.
\item $\Delta(s, a) = \{ \delta_u(s, a), \merge(\delta_g'(s, a)) \}$
  if $s \in S^u$ and $\delta_g'(s, a)$ is not doomed.
\item $\Delta(s, a) = \{ \merge(\delta_d'(s, a)) \}$ if $s \in S^d$
  and $\delta_d'(s, a)$ is not doomed.
\end{itemize}



Before proving the correctness of the improved \slice construction, we
define an-\break other merge function $\merge_{i,j}$ that will be used in the
proof.
For a decorated slice $s$, $\merge_{i,j}(s)$ is a slice
obtained from $s$ by merging as many as possible and at most $j$
consecutive mergible nodes starting from the $i$-th pair of
mergible nodes.
For example, if $s = (Q_0, 0)(Q_1, 0)(Q_2, 0)(Q_3, 0)(Q_4, 0)$, then
$\merge_{1,2}(s) = (Q_0 \cup Q_1, 0)(Q_2, 0)(Q_3, 0)(Q_4, 0)$ and
$\merge_{2,3}(s) = (Q_0, 0)(Q_1 \cup Q_2 \cup Q_3, 0)(Q_4, 0)$.
By this definition, $\merge(\merge_{i,j}(s)) = \merge(s)$ for any
$i$, $j$, and decorated slice $s$.

\begin{lem}
\label{lemma:slice-mergeij}
For a decorated slice $s \in S^d$ and a symbol $a \in \Sigma$, there
exist some $i$ and $j$ such that $\delta_d'(\merge_{1,2}(s), a) =
\merge_{i,j}(\delta_d'(s, a))$.
\end{lem}
\proof
Let $s = (Q_0,d_0)\cdots(Q_{n-1},d_{n-1})$ and $t =
(Q'_0,d'_0)\cdots(Q'_{2n-1},d'_{2n-1})$ be the $a$-successor of $s$
before removing empty nodes.
Assume $d_i = d_{i+1} \in \{0, *\}$, and $(Q_i,d_i)$ and
$(Q_{i+1},d_{i+1})$ are the first mergible pair of nodes in $s$.
By the decoration rules \rulename{D2} and \rulename{D4}, $d'_{2i} =
d'_{2i+1} = d'_{2i+2} = d'_{2i+3} \in \{0, *\}$.
Then, $\merge_{1,2}(s) = (Q_0,d_0)\cdots(Q_{i-1},d_{i-1})(Q_i \cup
Q_{i+1}, d_i)(Q_{i+2},d_{i+2})$\linebreak$\cdots(Q_{n-1},d_{n-1})$ and its
$a$-successor before removing empty nodes is
$(Q'_0,d'_0)\cdots(Q'_{2i-1},d'_{2i-1})$\linebreak$(Q_{2i} \cup Q_{2i+1} \cup
Q_{2i+2} \cup Q_{2i+3},
d_{2i})(Q'_{2i+4},d'_{2i+4})\cdots(Q'_{2n-1},d'_{2n-1})$, denote by
$t'$.
Since $d'_{2i} = d'_{2i+1} = d'_{2i+2} = d'_{2i+3} \in
\{0, *\}$, $(Q'_{2i}, d'_{2i})$, $(Q'_{2i+1}, d'_{2i+1})$, $(Q'_{2i+2},
d'_{2i+2})$, and $(Q'_{2i+3}, d'_{2i+3})$ are mergible.
Suppose $(Q'_{2i}, d'_{2i})$ and $(Q'_{2i+1}, d'_{2i+1})$ are the
$k$-th mergible pair of nodes.
Then, $\merge_{k, 4}(t) = t'$.
As $\delta'_d(s, a)$ and $\delta'_d(\merge_{1,2}(s), a)$ are
derived respectively from $t$ and $t'$ by removing empty nodes, we
can found some $i$ and $j$ ($0 \leq j \leq 4$) such that
$\delta_d'(\merge_{1,2}(s), a) = \merge_{i,j}(\delta_d'(s, a))$. \qed 

\begin{lem}
\label{lemma:slice-merge-one}
Let $s \in S^d$ be a decorated slice and $a \in \Sigma$ a symbol.
If $\delta_d'(s, a)$ is not doomed, then
$\Delta(\merge(s), a) = \{ \merge(\delta_d'(s, a)) \}$.
\end{lem}
\proof
We prove by induction on the number of mergible pairs in $s$.
The base case is that $s$ has no mergible pair, which implies that
$\merge(s) = s$.
Thus,
\begin{center}
\begin{tabular}{rcll}
$\Delta(\merge(s), a)$ & $=$ & $\{ \merge(\delta_d'(\merge(s), a)) \}$ & (by the definition of
$\Delta$) \\
& $=$ & $\{ \merge(\delta_d'(s, a)) \}$ & (by $\merge(s) = s$)
\end{tabular}
\end{center}

Assume the hypothesis holds for any slice that has $n$ mergible
pairs and consider a slice $s$ that has $n+1$ mergible pairs.
Since $\merge_{1,2}(s)$ has $n$ mergible pairs, we know that:
\[
\begin{array}{rcll}
\Delta(\merge(s), a) & = & \Delta(\merge(\merge_{1,2}(s)), a)  &
\mbox{(by the definition of $\merge_{i,j}$)} \\
& = & \{ \merge(\delta_d'(\merge_{1,2}(s), a)) \} & \mbox{(by the
  induction hypothesis)} \\ 
& = & \multicolumn{2}{l}{\{ \merge(\merge_{i,j}(\delta_d'(s, a))) \}
  \mbox{ for some $i$ and $j$}} \\ 
& & & \mbox{(by Lemma \ref{lemma:slice-mergeij})} \\
& = & \{ \merge(\delta_d'(s, a)) \} & \mbox{(by the definition of
  $\merge_{i,j}$)\rlap{\hbox to 25 pt{\hfill\qEd}}} 
\end{array}
\]

\begin{thm}
Given an NBW $A = (\Sigma, Q, q_0, \delta, \f)$, the improved \slice
construction produces an NBW $A' = (\Sigma, S, s_0, \Delta, \g)$ with $L(A') =
\overline{L(A)}$.
\end{thm}
\proof
We first prove that if a word $w = a_0 a_1 \cdots$ is rejected by $A$,
$w$ is accepted by $A'$.
Let $T = s_0 s_1 \cdots$ be the reduced split tree of $A$ on $w$ where
$s_{j+1} = \delta_u(s_j, a_j)$ for all $j$.
\begin{itemize}
\item Case 1: There is no run of $A$ on $w$.
  Then, there exists some $i$ such that $s_j = \bot$ for all $j \geq
  i$.
  By the construction of the improved \slice, $s_0 s_1 \cdots s_{i-1}
  \bot^\omega $ is an accepting run of $w$ on $A'$.
  Thus, $w$ is accepted by $A'$.
\item Case 2: There is at least one run of $A$ on $w$.
  Since $w$ is rejected by $A$, by Lemma \ref{lemma:slice-cutoff}, there
  exists some cutoff $i$ such that for all $j \geq i$, all accepting
  states of $A$ in $s_j$ belong to finite branches of $T$.
  Then, we can construct a sequence of slices $s_0 s_1 \cdots s_{i-1}
  t_i t_{i+1} \cdots$ where $t_j \in S^d$ such that
  \begin{itemize}
  \item $t_i = \delta_g'(s_{i-1}, a_{i-1})$,
  \item $t_{j+1} = \delta_d'(t_j, a_j)$ for $j \geq i$, and
  \item $\projq{t_j} = s_j$ for $j \geq i$.
  \end{itemize}
  As there is at least one run of $A$ on $w$, $t_j$ is not doomed for
  all $j \geq i$.
  By Lemma \ref{lemma:slice-merge-one} and the construction of the
  improved \slice, we can find a run $\rho = s_0 s_1 \cdots s_{i-1}
  u_i u_{i+1} \cdots$ of $A'$ on $w$ where $u_i \in \Delta(s_{i-1},
  a_{i-1})$, and for $j \geq i$, $u_{j+1} \in \Delta(u_j, a_j)$ and
  $u_j = \merge(t_j)$.
  Since all accepting states of $A$ in $s_j$ for $j \geq i$ belong to
  finite branches of $T$ and these states are decorated by either $0$
  or $*$ in both $t_j$ and $u_j$, we can find infinitely many reset
  slices in $\rho$ by the decoration rules.
  Thus, $\rho$ is accepting and $w$ is accepted by $A'$.
\end{itemize}

We then prove that if a word $w = a_0 a_1 \cdots$ is accepted by $A'$,
$w$ is rejected by $A$.
Let $\rho = s_0 s_1 \cdots$ be an accepting run of $A'$ on $w$.
\begin{itemize}
\item Case 1: $\bot \in \rho$.
  In this case, there is some $i$ such that $s_j = \bot$ for all $j
  \geq i$.
  Thus, there is no run of $A$ on $w$ and $w$ is rejected by $A$. 
\item Case 2: $\bot \not\in \rho$.
  Let $T = t_0 t_1 \cdots$ be the reduced split tree of $A$ on $w$. 
  Assume $s_i$ is the first decorated slice in $\rho$.
  Then, $s_j = t_j$ for $j < i$ and $s_j$ is not doomed for $j \geq
  i$.
  By Lemma~\ref{lemma:slice-merge-one} and the construction of the
  improved \slice, there is a sequence $\rho' = s_0 s_1 \cdots s_{i-1}
  u_i u_{i+1} \cdots$ where $u_j \in S^d$ for $j \geq i$ such that
  \begin{itemize}
  \item $u_i = \delta_g'(s_{i-1}, a_{i-1})$,
  \item $u_{j+1} = \delta_d'(u_j, a_j)$ for $j \geq i$, and
  \item $\merge(u_j) = s_j$ and $\projq{u_j} = t_j$ for $j \geq i$.
  \end{itemize}
  Since $\rho$ is accepting, there are infinitely many reset slices in
  $\rho$ as well as in $\rho'$.
  Based on the construction of the improved \slice, all accepting
  states of $A$ are decorated by either $0$ or $*$ in $\rho$,
  the decoration of $0$-nodes and $*$-nodes remains unchanged before
  the next reset slice, and the decoration of $*$-nodes becomes $0$
  after a reset slice.
  Thus, after $u_i$ in $\rho$, all these accepting states belong
  to finite branches.
  Since the $\merge$ function does not change any deocration, all
  these accepting states belong to finite branches after $u_i$ in
  $\rho'$.
  As $\rho'$ and $T$ only differ in decorations, all the accepting
  states of $A$ belong to finite branches after the $i$-th level in
  $T$.
  Hence, $w$ is rejected by $A$ according to
  Lemma \ref{lemma:slice-cutoff}. \qed
\end{itemize}
\hide{
We then prove that if a word $w$ is accepted by $A'$, $w$ is rejected by $A$.
Let $\rho = s_0 s_1 \cdots$ be an accepting run of $A'$ on $w$.
Based on the construction, there exists some $i$ such that for all $j \geq i$,
$s_j$ is decorated and all descendants of left nodes after $s_i$ are
decorated by either $0$ or $*$.
Since $\rho$ is accepting, it passes reset slices infinitely often, which
implies that (1) all $0$-nodes appear between two adjacent reset slices will
die out before the second reset slice and (2) all $*$-nodes appear between
two adjacent reset slices will die out before the next reset slice after the
second one.
Thus after $s_i$ in $\rho$, all left (accepting) nodes are on finite branches
even if the heuristic \texttt{+M} is applied.
This implies that the reduced split tree of $w$ on $A$ has no left-recurring
branch and hence $w$ is rejected by $A$.
}

\section{Experimental Results}\label{sec:experiment}

The heuristics proposed in Section~\ref{sec:optimization} were
implemented in GOAL.
For \ramsey, there is a naive optimization which minimizes the classic
finite automata before composing the NBWs to construct the
complement\footnote{The optimization heuristics for the Ramsey-based
  constructions proposed by Breuers
  \textit{et al.}~\cite{breuers:improved} were published after
  we had performed the experiments.
  Although their implementation and ours are not directly comparable,
  the average size of the complements produced by our \pitermanase
  construction without preminimization among the 10,980 
  finished tasks is 139.18 states (37.55 states after removing dead
  states).
  The improved Ramsey-based construction in~\cite{breuers:improved}
  finished 10,839 tasks with an average size of 361.09 states (328.97
  states after removing dead states).
  The maximal size of the complements is 5,238 states by \pitermanase
  and is 337,464 by the improved Ramsey-based construction.}.
This optimizaiton, refered to as $\texttt{+m}$, was also implemented
in GOAL.
We used the same 11,000 automata as in Section \ref{sec:comparison}  as
the test bench.
The results showing the improvement made by the heuristics are
summarized in Table~\ref{exp:opt1} where the Ratio columns are ratios
with respect to the original construction and the other columns have
the same meanings as in Section~\ref{sec:comparison}. 

\begin{table}[htb]
\caption{A comparison of each construction with its improved
  versions\label{exp:opt1}}
\small
\begin{center}
\begin{tabular}{|r|r|r|c|rr|rr|r|}
\hline \multicolumn{1}{|c|}{Constructions} & \multicolumn{1}{|c|}{$T$} &
\multicolumn{1}{|c|}{$M$} & Eff. Samples & $S_R$ & (Ratio) & $S_L$ & (Ratio) &
\multicolumn{1}{|c|}{$S_L/S_R$} \\ \hline\hline 
\ramsey   & 4,564 & 36 & 6,388 & 594.68 & (1.00) & 22.78 & (1.00) & 0.04 \\
\ramseya  & 4,557 & 33 &       & 595.59 & (1.00) & 22.54 & (0.99) & 0.04 \\
\ramseym  & 3,126 &  2 &       & 372.08 & (0.63) & 11.19 & (0.49) & 0.03 \\
\ramseyam & 3,119 &  2 &       & 371.06 & (0.62) & 11.02 & (0.48) & 0.03 \\ \hline\hline

\SP    &  5 & 0 & 10,977 & 256.25 & (1.00) & 58.72 & (1.00) & 0.23 \\
\SPa   &  5 & 0 &        & 228.40 & (0.89) & 54.33 & (0.93) & 0.24 \\
\SPs   & 12 & 9 &        & 179.82 & (0.70) & 47.35 & (0.81) & 0.26 \\
\SPe   & 11 & 0 &        & 194.95 & (0.76) & 45.47 & (0.77) & 0.23 \\
\SPase & 13 & 7 &        & 138.97 & (0.54) & 37.47 & (0.64) & 0.27 \\ \hline\hline

\rank  & 5,303 & 0 & 5,697 & 569.51 & (1.00) & 33.96 & (1.00) & 0.06 \\
\ranka & 3,927 & 0 &       & 181.05 & (0.32) & 28.41 & (0.84) & 0.16 \\ \hline\hline

\slice     & 3,131 & 3,213 & 4,514 & 1,088.72 & (1.00) &  70.67 & (1.00) & 0.06 \\
\slicea    & 2,611 & 2,402 &       &   684.07 & (0.63) &  64.94 & (0.92) & 0.09 \\
\sliced    & 1,119 &     0 &       &   276.11 & (0.25) & 117.32 & (1.66) & 0.42 \\
\slicer    & 3,081 & 3,250 &       & 1,028.42 & (0.94) &  49.58 & (0.70) & 0.05 \\
\slicem    & 2,813 & 3,360 &       &   978.01 & (0.90) &  57.85 & (0.82) & 0.06 \\
\sliceadrm &   228 &     0 &       &   102.57 & (0.09) &  36.11 & (0.51) & 0.35 \\ \hline
\end{tabular}
\end{center}
\end{table}

The experimental results in Table~\ref{exp:opt1} show that
(1) the heuristic \texttt{+m} can reduce states down to around one half for
\ramsey,
(2) \pitermanase has 15 more unfinished tasks but creates just around one
half of reachable states and live states,
(3) the improvement made by \texttt{+A} is limited for \ramsey, \piterman, and
\slice but substantial for \rank in helping finish 1,376 more tasks and
avoid the creation of around $2/3$ dead states,
(4) the heuristic \texttt{+D} is quite useful in reducing the reachable states
down to $1/4$ for \slice but produces more live states, and
(5) \sliceadrm finishes 6,116 more tasks and significantly reduces the reachable
states to $1/10$ and live states to one half. 

We also compared the four constructions with all optimization
heuristics in Section~\ref{sec:optimization} based on 4,851 effective
samples and list the results on the top of Table~\ref{exp:opt2_size15}. 
The table shows that \pitermanase still outperforms the other three in
the average state size and in running time.
Table~\ref{exp:opt2_size15} also shows the following changes made by
our heuristics in the comparison: (1) \pitermanase outperforms \ranka
in the number of smallest complements after pruning dead states, and
(2) \sliceadrm creates fewer reachable states than \ranka in average,
and finishes more tasks than \ranka and \ramseyam.

Same as in Section~\ref{sec:comparison}, we also compared the four
improved constructions with preminimization.
The results are summarized on the bottom of
Table~\ref{exp:opt2_size15}.
Similarly, the preminimization does improve \ramsey, \rank, and \slice
more than \piterman in the complementation but does not close too much
the gap between them in the comparison.

\begin{table}[t]
\caption{A comparison of the four improved complementation
  constructions based on $\set{A}_{15}$ without
  and with preminimization\label{exp:opt2_size15}}
\small
\begin{center}
\begin{tabular}{|r|r|r|c|rr|rr|r|}
\hline \multicolumn{1}{|c|}{Constructions} & \multicolumn{1}{|c|}{$T$} &
\multicolumn{1}{|c|}{$M$} & Eff. Samples & $S_R$ & (Win) & $S_L$ & (Win) &
\multicolumn{1}{|c|}{$S_L/S_R$} \\ \hline\hline
\multicolumn{9}{|l|}{$\set{A}_{15}$ (without preminimization)} \\ \hline
\ramseyam    & 3,119 & 2 & 4,851 & 666.86 &     (0.0) & 251.53 &   (895.75) & 0.38 \\
\SPase &    13 & 7 &       &  23.88 & (4,772.5) &   5.77 & (1,675.42) & 0.24 \\
\ranka       & 3,927 & 0 &       & 384.35 &    (20.0) &  11.38 & (1,138.42) & 0.03 \\ 
\sliceadrm   &   228 & 0 &       & 185.02 &    (58.5) &   9.65 & (1,141.42) & 0.05 \\ \hline\hline
\multicolumn{9}{|l|}{$\set{A}_{15}$ (with preminimization)} \\ \hline
\ramseypam    & 2,825 & 6 & 5,618 & 479.99 &    (17.50) & 225.08 & (1,031.25) & 0.47 \\
\SPpase &    12 & 7 &       &  19.47 & (3,820.67) &   5.89 & (1,698.75) & 0.30 \\
\rankpa       & 3,383 & 0 &       & 232.63 &   (875.67) &  10.68 & (1,476.75) & 0.05 \\ 
\slicepadrm   &   216 & 0 &       & 135.74 &   (904.17) &   9.12 & (1,411.25) & 0.07 \\ \hline
\end{tabular}
\end{center}
\end{table}

The comparisons of the four improved constructions based on the
nonuniversal automata in $\set{A}_{15}$ without and with
preminimization are summarized in
Table~\ref{exp:opt2_size15_nonuniv}.
These comparisons based on the nonuniversal automata are quite
consistent to those based all the 11,000 automata.
Additional comparisons of the four improved constructions based on
$\set{A}_{10}$ and $\set{A}_{20}$ can be found in the appendix.

\begin{table}[htb]
\caption{A comparison of the four improved complementation
  constructions based on the nonuniversal automata in
  $\set{A}_{15}$ without and with
  preminimization\label{exp:opt2_size15_nonuniv}}
\small
\begin{center}
\begin{tabular}{|r|c|rr|rr|r|}
\hline \multicolumn{1}{|c|}{Constructions} & Eff. Samples & $S_{R}$ &
(Win) & $S_{L}$ & (Win) & \multicolumn{1}{|c|}{$S_L/S_R$} \\
\hline\hline
\multicolumn{7}{|l|}{$\set{A}_{15}$ (without preminimization)} \\ \hline
\ramseyam    & 1,270 & 1,310.06 &    (0.0) & 957.96 &    (0.50) & 0.731 \\
\SPase &       &    39.90 & (1191.5) &  19.21 &  (780.17) & 0.482 \\
\ranka       &       &   314.22 &   (20.0) &  40.64 &  (243.17) & 0.129 \\
\sliceadrm   &       &   186.90 &   (58.5) &  34.04 &  (246.17) & 0.182 \\ \hline
\multicolumn{7}{|l|}{$\set{A}_{15}$ (with preminimization)} \\ \hline
\ramseypam    & 1,495 & 1,125.22 &    (1.0) & 843.07 &    (0.5) & 0.749 \\
\SPpase &       &    38.71 & (1110.5) &  19.39 &  (668.0) & 0.501 \\
\rankpa       &       &   260.28 &  (177.0) &  37.39 &  (446.0) & 0.144 \\
\slicepadrm   &       &   163.29 &  (206.5) &  31.47 &  (380.5) & 0.193 \\ \hline
\end{tabular}
\end{center}
\end{table}

\section{Conclusion}\label{sec:conclusion}

We reviewed the state of \buchi complementation and examined the
performance of the four complementation approaches by experiments
with our implementations in GOAL and three test sets of 11,000
automata.
The experimental results showed that the determinization-based
approach performs better than the other three in average.
In our implementations, the Ramsey-based approach is not competitive
in complementation though it is competitive in universality and
containment testing.

We also proposed various optimization heuristics for three of the
approaches and performed an experiment with one of the test sets to
show the improvement.
The experimental results also showed that our heuristics substantially
improve \piterman and \slice in creating far fewer states.
\rank and especially \slice can finish more complementation tasks with
our heuristics.

As the experimental results showed, the nondeterministic constructions
\rank and \slice produced many more dead states of complements.
One reason is that there are many nondeterministic choices (rank
functions in the rank-based approach and decorations in the
slice-based approach) but only few of them are correct.
While Friedgut \textit{et al.} proposed tight ranking
in~\cite{friedgut:buchi} to reduce the number of ranking functions for
the rank-based approach, we proposed the heuristic of deterministic
decoration to alleviate this problem for the slice-based approach.
However, the improved constructions \ranka and \sliceadrm still
produced many more dead states compared to \pitermanase in our
experiments.
There may be other opportunities to improve the rank-based and the
slice-based constructions further.


\bibliographystyle{plain}
\bibliography{references}

\appendix

\section{Full Experimental Results}

This section includes the full experimental results that we performed
to compare the four representative complementation constructions and
compare their improved verions.
In the following, we first recall the settings of our experiments and
then describe the comparisons based on the experimental results.

Let the parameter $n$ be $10$, $15$, or $20$, which denotes a size of
states.
For each $n$, we randomly generated 11,000 automata with an
alphabet of size 2 and states of size $n$ as a test set.
Among the 11,000 automata of state size $n$, denoted by $\set{A}_{n}$,
100 automata are generated from each combination of 11 transition
densities (from 1.0 to 3.0) and 10 acceptance densities (from 0.1 to
1.0).
For every generated automaton $A = (\Sigma, Q, q_0, \delta, \f)$ with
a given state size $n$, symbol $a \in \Sigma$, transition density $r$,
and acceptance density $f$, we made $q \in \delta(p, a)$ for $\lceil
rn \rceil$ pairs of states $(p, q) \in Q^2$ uniformly chosen at random
and added $\lceil fn \rceil$ states to $\f$ uniformly at random.
Our parameters were chosen to generate a large set of complementation
problems, ranging from easy to hard.

We chose four representative complementation constructions, namely
\ramsey \cite{sistla:complementation},
\piterman~\cite{piterman:buchi}, \rank~\cite{schewe:buchi}, and
\slice~\cite{vardi:automata07}, each of which is considered the most
efficient construction in its respective approach.
These constructions were implemented in
the GOAL tool~\cite{tsay:goal,tsai:goal}.
The experiment was performed on a cluster at Rice University
(http://rcsg.rice.edu/sugar/int/) with GOAL based on the three
generated test sets $\set{A}_{10}$, $\set{A}_{15}$, and
$\set{A}_{20}$.
For each complementation task, we allocated one 2.83-GHz CPU and 1 GB
of memory.
The timeout of a complementation task was 10 minutes.

\subsection{Comparisons of Basic Constructions\label{sec:appendix:comparison}}

\begin{table}[htb]
\caption{A comparison of the four representative
  constructions based on $\set{A}_{10}$, $\set{A}_{15}$, and
  $\set{A}_{20}$\label{exp:orig_101520}}
\small
\begin{center}
\begin{tabular}{|r|r|r|c|rr|rr|r|}
\hline \multicolumn{1}{|c|}{Constructions} & \multicolumn{1}{|c|}{$T$} &
\multicolumn{1}{|c|}{$M$} & Eff. Samples & $S_{R}$ & (Win) & $S_{L}$ & (Win) &
\multicolumn{1}{|c|}{$S_L/S_R$} \\ \hline\hline
\multicolumn{9}{|l|}{$\set{A}_{10}$} \\ \hline
\ramsey   & 2,944 &  81 & 5,056 & 406.55 &   (70.5) & 50.06 & (1085.75) & 0.123 \\
\SP       &     0 &   0 &       &  39.05 & (3834.0) &  3.50 & (1312.92) & 0.090 \\
\rank     & 2,667 &   0 &       & 462.04 & (1150.5) &  7.63 & (1520.92) & 0.017 \\
\slice    & 1,191 & 482 &       & 913.57 &    (1.0) &  4.92 & (1136.42) & 0.005 \\ \hline\hline
\multicolumn{9}{|l|}{$\set{A}_{15}$} \\ \hline
\ramsey   & 4,564 &    36 & 2,259 & 513.85 &     (0) & 30.82 & (522.50) & 0.060 \\
\SP       &     5 &     0 &       &  45.26 & (1,843) &  2.27 & (556.67) & 0.050 \\
\rank     & 5,303 &     0 &       & 260.41 &   (415) &  2.79 & (649.17) & 0.011 \\
\slice    & 3,131 & 3,213 &       & 790.92 &     (1) &  3.03 & (530.67) & 0.004 \\ \hline\hline
\multicolumn{9}{|l|}{$\set{A}_{20}$} \\ \hline
\ramsey   & 5,588 &   240 & 1,390 & 549.24 &     (0) & 17.38 & (335.5) & 0.032 \\
\SP       &    53 &     0 &       &  57.41 & (1,101) &  1.88 & (348.5) & 0.033 \\
\rank     & 6,784 &     0 &       & 290.17 &   (289) &  2.17 & (370.5) & 0.007 \\
\slice    & 3,647 & 4,224 &       & 736.94 &     (0) &  2.42 & (335.5) & 0.003 \\ \hline
\end{tabular}
\end{center}
\end{table}

The comparisons of the four representative constructions based on the
three test sets are summarized in Table~\ref{exp:orig_101520} where
$T$ is the total number of timed-out tasks and $M$ is the total number
of tasks that run out of memory.
The column $S_R$ is the average number of reachable states, while
$S_L$ is the average number of live states, of the complements.
The column Eff. Samples denotes the total number of effective samples
where both $S_R$ and $S_L$ are calculated.
There are 5056, 2259, and 1390 effective samples respectively in
$\set{A}_{10}$, $\set{A}_{15}$, and $\set{A}_{20}$.
Among the effective samples of each test set, around 90\% of the
automata are universal.
The Win column of a construction in $S_R$ (resp., $S_L$) denotes the
fractional share of effective samples where the construction wins
w.r.t. $S_R$ (resp., $S_L$).
More detailed definition of effective samples and win shares can be
found in Section~\ref{sec:comparison}.

The number of unfinished tasks by a construction is the sum of $T$ and
$M$.
Compared to \piterman that has less than 0.5\% of the unfinished tasks
in all the test sets, each of \ramsey, \rank, and \slice has more than
15\% in $\set{A}_{10}$, more than 40\% in $\set{A}_{15}$, and more
than 50\% in $\set{A}_{20}$.

The columns $S_R$ and $S_L$ show that \piterman is the best in average
state size.
The low $S_L/S_R$ ratio shows that \rank and \slice create more dead 
states that can be easily pruned off.
Although \rank generates more dead states than \piterman, the Win
column of $S_L$ shows that \rank produces more complements that are
the smallest after pruning dead states.

\rank and \slice become much closer to \piterman in $S_L$ because
around 90\% of the effective samples are universal automata,
whose complements have no live states.
If we only consider nonuniversal automata, the gaps between \piterman
and \rank, and \piterman and \slice in $S_L$ become larger as
shown in Table~\ref{exp:orig_nonuniv_101520}.
This case also happens in the the Win column of $S_L$ between
\piterman and \ramsey, and \piterman and \slice.

\begin{table}[htb]
\caption{A comparison of the four representative
  constructions based on the nonuniversal automata in
  $\set{A}_{10}$, $\set{A}_{15}$, and
  $\set{A}_{20}$\label{exp:orig_nonuniv_101520}}
\small
\begin{center}
\begin{tabular}{|r|c|rr|rr|r|}
\hline \multicolumn{1}{|c|}{Constructions} & Eff. Samples & $S_{R}$ &
(Win) & $S_{L}$ & (Win) & \multicolumn{1}{|c|}{$S_L/S_R$} \\
\hline\hline
\multicolumn{7}{|l|}{$\set{A}_{10}$} \\ \hline
\ramsey   & 713 & 1,600.64 &     (0) & 348.88 &      (0) & 0.644 \\
\SP       &     &    47.23 & (493.5) &  18.74 & (227.17) & 0.397 \\
\rank     &     &   437.58 & (218.5) &  48.03 & (435.17) & 0.110 \\
\slice    &     &   686.04 &     (1) &  28.77 &  (50.67) & 0.042 \\ \hline
\multicolumn{7}{|l|}{$\set{A}_{15}$} \\ \hline
\ramsey   & 171 & 1,892.37 &     (0) & 397.10 &      (0) & 0.210 \\
\SP       &     &    36.38 & (102.5) &  17.77 &  (34.67) & 0.488 \\
\rank     &     &   156.63 &  (67.5) &  24.61 & (127.67) & 0.157 \\
\slice    &     &   422.88 &     (1) &  27.75 &   (8.67) & 0.066 \\ \hline
\multicolumn{7}{|l|}{$\set{A}_{20}$} \\ \hline
\ramsey   & 48 & 2,052.02 &  (0) & 475.27 &  (0) & 0.232 \\
\SP       &    &    41.13 & (30) &  26.58 & (13) & 0.646 \\
\rank     &    &   165.88 & (18) &  34.75 & (35) & 0.209 \\
\slice    &    &   206.90 &  (0) &  42.02 &  (0) & 0.203 \\ \hline
\end{tabular}
\end{center}
\end{table}


As the heuristic of preminimization applied to the input automata,
denoted by \texttt{+P}, is considered to help the nondeterministic
constructions more than the deterministic one, we also compare the
four constructions with preminimization and summarize the results in
Table~\ref{exp:orig_101520_opt}.
The results based on nonuniversal automata are summarized in
Table~\ref{exp:orig_nonuniv_101520_opt}.
We only applied the preminimization implemented in GOAL, namely the
simplification by simulation in~\cite{somenzi:efficient}. 
According to our experimental results, the preminimization does
improve \ramsey, \rank, and \slice more than \piterman in the
complementation but does not close too much the gap between them in
the comparison, though there are other preminimization techniques that
we did not apply in the experiment.

\begin{table}[htb]
\caption{A comparison of the four representative
  constructions based on $\set{A}_{10}$, $\set{A}_{15}$, and
  $\set{A}_{20}$ with preminimization\label{exp:orig_101520_opt}}
\small
\begin{center}
\begin{tabular}{|r|r|r|c|rr|rr|r|}
\hline \multicolumn{1}{|c|}{Constructions} & \multicolumn{1}{|c|}{$T$} &
\multicolumn{1}{|c|}{$M$} & Eff. Samples & $S_{R}$ & (Win) & $S_{L}$ & (Win) &
\multicolumn{1}{|c|}{$S_L/S_R$} \\ \hline\hline
\multicolumn{9}{|l|}{$\set{A}_{10}$} \\ \hline
\ramseyp   & 2,524 &  85 & 6,142 & 273.18 &   (405.83) & 46.75 & (1,273.83) & 0.171 \\
\SPp       &     0 &   0 &       &  22.01 & (3,452.83) &  3.46 & (1,502.67) & 0.157 \\
\rankp     & 2,006 &   0 &       & 278.00 & (2,244.33) &  8.92 & (2,009.33) & 0.032 \\
\slicep    & 1,064 & 300 &       & 531.49 &    (39.00) &  4.98 & (1,356.17) & 0.010 \\ \hline\hline
\multicolumn{9}{|l|}{$\set{A}_{15}$} \\ \hline
\ramseyp   & 4,190 &    41 & 3,522 & 193.72 &   (247) & 26.82 &   (776.25) & 0.218 \\
\SPp       &     4 &     0 &       &  11.08 & (1,712) &  2.31 &   (817.42) & 0.572 \\
\rankp     & 4,316 &     0 &       &  56.60 & (1,546) &  2.37 & (1,126.42) & 0.160 \\
\slicep    & 2,908 & 2,435 &       & 199.52 &    (17) &  2.94 &   (801.92) & 0.092 \\ \hline
\multicolumn{9}{|l|}{$\set{A}_{20}$} \\ \hline
\ramseyp   & 5,334 &   185 & 2,623 & 133.35 &   (173.67) & 17.81 & (599.25) & 0.134 \\
\SPp       &    44 &     0 &       &   8.10 & (1,235.67) &  1.82 & (609.25) & 0.224 \\
\rankp     & 5,758 &     0 &       &  35.90 & (1,198.17) &  1.66 & (808.75) & 0.046 \\
\slicep    & 3,343 & 3,559 &       & 100.78 &     (15.5) &  2.18 & (605.75) & 0.021 \\ \hline
\end{tabular}
\end{center}
\end{table}

\begin{table}[htb]
\caption{A comparison of the four representative
  constructions based on the nonuniversal automata in $\set{A}_{10}$,
  $\set{A}_{15}$, and $\set{A}_{20}$ with
  preminimization\label{exp:orig_nonuniv_101520_opt}}
\small
\begin{center}
\begin{tabular}{|r|c|rr|rr|r|}
\hline \multicolumn{1}{|c|}{Constructions} & Eff. Samples & $S_{R}$ &
(Win) & $S_{L}$ & (Win) & \multicolumn{1}{|c|}{$S_L/S_R$} \\
\hline\hline
\multicolumn{7}{|l|}{$\set{A}_{10}$} \\ \hline
\ramseyp   & 1,049 & 1,160.55 &   (0.0) & 268.88 &   (0.58) & 0.232 \\
\SPp       &       &    36.21 & (481.5) &  15.41 & (229.42) & 0.426 \\
\rankp     &       &   323.76 & (565.5) &  47.37 & (736.08) & 0.146 \\
\slicep    &       &   450.58 &   (2.0) &  24.27 &  (82.92) & 0.054 \\ \hline
\multicolumn{7}{|l|}{$\set{A}_{15}$} \\ \hline
\ramseyp   & 418 & 1,000.89 &   (0.0) & 218.55 &   (0.25) & 0.218 \\
\SPp       &     &    21.02 & (116.0) &  12.02 &  (41.42) & 0.572 \\
\rankp     &     &    77.97 & (300.5) &  12.51 & (350.42) & 0.160 \\
\slicep    &     &   189.90 &   (1.5) &  17.39 &  (25.92) & 0.092 \\ \hline
\multicolumn{7}{|l|}{$\set{A}_{20}$} \\ \hline
\ramseyp   & 226 & 879.32 &   (0) & 196.14 &   (0) & 0.223 \\
\SPp       &     &  15.51 &  (51) &  10.50 &  (11) & 0.677 \\
\rankp     &     &  23.81 & (179) &   8.70 & (216) & 0.365 \\
\slicep    &     &  68.17 &   (0) &  14.69 &  (12) & 0.216 \\ \hline
\end{tabular}
\end{center}
\end{table}

In summary, the experimental results show that (1) \piterman is the
best in average state size and in the number of finished tasks, (2)
\ramsey is not competitive in complementation even though it is
competitive in universality and containment testing as shown
in~\cite{abdulla:when,fogarty:buchi,fogarty:efficient}, and (3) except
in $\set{A}_{10}$, \slice has the most unfinished tasks (even more
than \ramsey) and, compared to \piterman and \rank, produces many more
states.

\subsection{Comparisons of Improved Constructions}

We also compared the four constructions with all optimization
heuristics in Section~\ref{sec:optimization} and one for \ramsey in
Section~\ref{sec:experiment} based on 7963, 4851, and 2951 effective
samples respectively in $\set{A}_{10}$, $\set{A}_{15}$, and
$\set{A}_{20}$.
Recall that the following heuristics were proposed in this paper:
simplifying DPWs by simulation (\texttt{+S}) and merging equivalent 
states (\texttt{+E}) for \piterman; maximizing the \buchi acceptance
set (\texttt{+A}) for \rank; deterministic decoration (\texttt{+D}),
reducing transitions (\texttt{+R}), and merging adjacent nodes
(\texttt{+M}) for \slice.
The heuristic for \ramsey is the simplification of intermediate
classic finite automata on finite words (\texttt{+m}).
The comparisons are summarized in Table~\ref{exp:opt2_101520}, which 
shows that \pitermanase still outperforms the other three in the
average state size and in running time.
Table~\ref{exp:opt2_101520} also shows the following changes made by
our heuristics in the comparison: (1) \pitermanase outperforms \ranka
in the number of smallest complements after pruning dead states, and
(2) \sliceadrm creates fewer reachable states than \ranka in average,
and finishes more tasks than \ranka and \ramseyam.

\begin{table}[htb]
\caption{A comparison of the four improved constructions based on
  $\set{A}_{10}$, $\set{A}_{15}$, and
  $\set{A}_{20}$\label{exp:opt2_101520}}
\small
\begin{center}
\begin{tabular}{|r|r|r|c|rr|rr|r|}
\hline \multicolumn{1}{|c|}{Constructions} & \multicolumn{1}{|c|}{$T$} &
\multicolumn{1}{|c|}{$M$} & Eff. Samples & $S_R$ & (Win) & $S_L$ & (Win) &
\multicolumn{1}{|c|}{$S_L/S_R$} \\ \hline\hline
\multicolumn{9}{|l|}{$\set{A}_{10}$} \\ \hline
\ramseyam    &   924 & 3 & 7,963 & 461.56 &     (9.50) & 257.21 & (1,323.75) & 0.56 \\
\SPase       &     0 & 0 &       &  26.96 & (7,701.83) &   7.42 & (3,062.75) & 0.28 \\
\ranka       & 2,285 & 0 &       & 438.66 &    (71.83) &  23.28 & (1,818.75) & 0.05 \\ 
\sliceadrm   &     2 & 0 &       & 115.79 &   (179.83) &  12.76 & (1,757.75) & 0.11 \\ \hline\hline
\multicolumn{9}{|l|}{$\set{A}_{15}$} \\ \hline
\ramseyam    & 3,119 & 2 & 4,851 & 666.86 &     (0.0) & 251.53 &   (895.75) & 0.38 \\
\SPase       &    13 & 7 &       &  23.88 & (4,772.5) &   5.77 & (1,675.42) & 0.24 \\
\ranka       & 3,927 & 0 &       & 384.35 &    (20.0) &  11.38 & (1,138.42) & 0.03 \\ 
\sliceadrm   &   228 & 0 &       & 185.02 &    (58.5) &   9.65 & (1,141.42) & 0.05 \\ \hline\hline
\multicolumn{9}{|l|}{$\set{A}_{20}$} \\ \hline
\ramseyam    & 5,009 &  14 & 2,951 & 618.07 &     (0.00) & 125.76 & (618.75) & 0.20 \\
\SPase       &    83 & 133 &       &  18.81 & (2,929.67) &   3.81 & (894.75) & 0.20 \\
\ranka       & 4,955 &   0 &       & 427.75 &     (5.17) &   8.41 & (717.25) & 0.02 \\ 
\sliceadrm   & 1,220 &   0 &       & 213.76 &    (16.17) &   5.96 & (720.25) & 0.03 \\ \hline
\end{tabular}
\end{center}
\end{table}

Same as in Section~\ref{sec:appendix:comparison}, we also compared the
four improved constructions with preminimization.
The results are summarized in Table~\ref{exp:opt3_101520}.
Similarly, the preminimization does improve \ramsey, \rank, and \slice
more than \piterman in the complementation but does not close too much
the gap between them in the comparison.

\begin{table}[hbt]
\caption{A comparison of the four improved constructions based on
  $\set{A}_{10}$, $\set{A}_{15}$, and
  $\set{A}_{20}$ with preminimization\label{exp:opt3_101520}}
\small
\begin{center}
\begin{tabular}{|r|r|r|c|rr|rr|r|}
\hline \multicolumn{1}{|c|}{Constructions} & \multicolumn{1}{|c|}{$T$} &
\multicolumn{1}{|c|}{$M$} & Eff. Samples & $S_R$ & (Win) & $S_L$ & (Win) &
\multicolumn{1}{|c|}{$S_L/S_R$} \\ \hline\hline
\multicolumn{9}{|l|}{$\set{A}_{10}$} \\ \hline
\ramseypam    &   813 & 6 & 8,565 & 355.43 &    (53.0) & 220.28 & (1,453.25) & 0.62 \\
\SPpase       &     0 & 0 &       &  22.25 & (6,032.0) &   6.97 & (2,794.25) & 0.31 \\
\rankpa       & 1,765 & 0 &       & 306.92 & (1,209.5) &  20.54 & (2,208.25) & 0.07 \\
\slicepadrm   &     2 & 0 &       &  89.70 & (1,270.5) &  11.40 & (2,109.25) & 0.13 \\ \hline\hline
\multicolumn{9}{|l|}{$\set{A}_{15}$} \\ \hline
\ramseypam    & 2,825 & 6 & 5,618 & 479.99 &    (17.50) & 225.08 & (1,031.25) & 0.47 \\
\SPpase       &    12 & 7 &       &  19.47 & (3,820.67) &   5.89 & (1,698.75) & 0.30 \\
\rankpa       & 3,383 & 0 &       & 232.63 &   (875.67) &  10.68 & (1,476.75) & 0.05 \\ 
\slicepadrm   &   216 & 0 &       & 135.74 &   (904.17) &   9.12 & (1,411.25) & 0.07 \\ \hline\hline
\multicolumn{9}{|l|}{$\set{A}_{20}$} \\ \hline
\ramseypam    & 4,647 &  15 & 3,741 & 390.30 &    (16.5) & 159.04 &   (762.75) & 0.41 \\
\SPpase       &   102 & 110 &       &  13.51 & (2,335.0) &   4.49 & (1,059.42) & 0.33 \\
\rankpa       & 4,422 &   0 &       & 208.18 &   (685.0) &   7.76 &   (964.92) & 0.04 \\
\slicepadrm   & 1,180 &   0 &       & 133.69 &   (704.5) &   6.66 &   (953.92) & 0.05 \\ \hline
\end{tabular}
\end{center}
\end{table}

The comparisons of the four improved constructions based on the
nonuniversal automata without and with preminimization are summarized
respectively in Table~\ref{exp:opt2_nonuniv_101520} and in
Table~\ref{exp:opt3_nonuniv_101520}.
These comparisons based on the nonuniversal automata are quite
consistent to those based all the automata.

\begin{table}
\caption{A comparison of the four improved constructions based on the
  nonuniversal automata in $\set{A}_{10}$, $\set{A}_{15}$, and
  $\set{A}_{20}$\label{exp:opt2_nonuniv_101520}}
\small
\begin{center}
\begin{tabular}{|r|c|rr|rr|r|}
\hline \multicolumn{1}{|c|}{Constructions} & Eff. Samples & $S_{R}$ &
(Win) & $S_{L}$ & (Win) & \multicolumn{1}{|c|}{$S_L/S_R$} \\
\hline\hline
\multicolumn{7}{|l|}{$\set{A}_{10}$} \\ \hline
\ramseyam    & 2,672 & 944.62 &        (1) & 764.56 &     (1) & 0.809 \\
\SPase       &       &  41.65 & (2,428.17) &  20.14 & (1,740) & 0.484 \\
\ranka       &       & 356.82 &    (67.67) &  67.39 &   (496) & 0.189 \\
\sliceadrm   &       & 116.78 &   (175.17) &  36.05 &   (435) & 0.309 \\ \hline
\multicolumn{7}{|l|}{$\set{A}_{15}$} \\ \hline
\ramseyam    & 1,270 & 1,310.06 &    (0.0) & 957.96 &    (0.50) & 0.731 \\
\SPase       &       &    39.90 & (1191.5) &  19.21 &  (780.17) & 0.482 \\
\ranka       &       &   314.22 &   (20.0) &  40.64 &  (243.17) & 0.129 \\
\sliceadrm   &       &   186.90 &   (58.5) &  34.04 &  (246.17) & 0.182 \\ \hline
\multicolumn{7}{|l|}{$\set{A}_{20}$} \\ \hline
\ramseyam    & 478 & 1,117.59 &   (0.00) & 772.57 &   (0) & 0.691 \\
\SPase       &     &    35.62 & (456.67) &  18.35 & (277) & 0.515 \\
\ranka       &     &   350.93 &   (5.17) &  46.74 &  (99) & 0.133 \\
\sliceadrm   &     &   219.55 &  (16.17) &  31.59 & (102) & 0.144 \\ \hline
\end{tabular}
\end{center}
\end{table}
\newpage

\begin{table}
\caption{A comparison of the four improved constructions based on the
  nonuniversal automata in $\set{A}_{10}$, $\set{A}_{15}$, and
  $\set{A}_{20}$ with preminimization\label{exp:opt3_nonuniv_101520}}
\small
\begin{center}
\begin{tabular}{|r|c|rr|rr|r|}
\hline \multicolumn{1}{|c|}{Constructions} & Eff. Samples & $S_{R}$ &
(Win) & $S_{L}$ & (Win) & \multicolumn{1}{|c|}{$S_L/S_R$} \\
\hline\hline
\multicolumn{7}{|l|}{$\set{A}_{10}$} \\ \hline
\ramseypam    & 2,752 & 845.03 &     (1.5) & 682.96 &     (0.5) & 0.808 \\
\SPpase       &       &  39.54 & (2,096.5) &  19.57 & (1,341.5) & 0.495 \\
\rankpa       &       & 305.25 &   (294.5) &  61.78 &   (755.5) & 0.202 \\
\slicepadrm   &       & 107.76 &   (361.5) &  33.35 &   (656.5) & 0.309 \\ \hline
\multicolumn{7}{|l|}{$\set{A}_{15}$} \\ \hline
\ramseypam    & 1,495 & 1,125.22 &    (1.0) & 843.07 &    (0.5) & 0.749 \\
\SPpase       &       &    38.71 & (1110.5) &  19.39 &  (668.0) & 0.501 \\
\rankpa       &       &   260.28 &  (177.0) &  37.39 &  (446.0) & 0.144 \\
\slicepadrm   &       &   163.29 &  (206.5) &  31.47 &  (380.5) & 0.193 \\ \hline
\multicolumn{7}{|l|}{$\set{A}_{20}$} \\ \hline
\ramseypam    & 692 & 1,084.23 &   (0.00) & 856.38 &   (0.00) & 0.790 \\
\SPpase       &     &    33.57 & (478.33) &  19.86 & (297.67) & 0.592 \\
\rankpa       &     &   224.91 &  (97.33) &  37.55 & (202.67) & 0.167 \\
\slicepadrm   &     &   165.83 & (116.33) &  31.61 & (191.67) & 0.191 \\ \hline
\end{tabular}
\end{center}
\end{table}
\relax
{\cW Here is some more text.}

\end{document}